\documentclass[letter,apj]{emulateapj-rtx4}

\pdfoutput=1

\usepackage{graphicx}
\usepackage{enumerate}
\usepackage{amssymb, amsmath, amsfonts}
\usepackage{gensymb}
\usepackage{mathtools}
\usepackage{natbib}
\usepackage{color}
\usepackage{hyperref}
\usepackage{ulem}
\usepackage{soul}
\usepackage{url}
\usepackage{xcolor}
\usepackage{xspace}
\usepackage{makecell} 

\usepackage{array}

\usepackage[referable]{threeparttablex}
\newcolumntype{L}{>{\centering\arraybackslash}m{3cm}}
\newcolumntype{C}{>{\centering\arraybackslash\tiny}m{3cm}}

\newcommand{\metric}{\Delta^2}
\newcommand{\finalscatter}{5 \cdot 10^{-3}}
\newcommand{\finalprecision}{10}

\shorttitle{Automatic Wavelength Calibration}
\shortauthors{Brandt \& Brandt et al.}

\begin{document}

\altaffiltext{1}{Department of Physics, University of California, Santa Barbara, Santa Barbara, CA 93106, USA}
\altaffiltext{2}{Las Cumbres Observatory, 6740 Cortona Drive, Suite 102, Goleta, CA 93117-5575, USA}
\altaffiltext{3}{NSF Graduate Research Fellow}

\title{Automatic \'Echelle Spectrograph Wavelength Calibration}
\author{
G.~Mirek Brandt\altaffilmark{1, 2, 3}, 
Timothy D.~Brandt\altaffilmark{1}, and
Curtis McCully\altaffilmark{2}
}

{Submitted to AJ}

\begin{abstract}
Time domain astronomy and the increasing number of exoplanet candidates call for reliable, robust, and automatic wavelength calibration. We present an algorithm for wavelength calibrating \'echelle spectrographs that uses order-by-order extracted spectra and a list of laboratory wavelengths. Our approach is fully automatic and does not need the pixel locations of certain spectral features with which to anchor the wavelength solution, nor the true order number of each diffraction order. We use spectral features that are duplicated in adjacent orders to establish the scale-invariant component of the wavelength solution. We then match the central wavelengths of spectral features to laboratory wavelengths to establish the scale and higher order components of the wavelength solution. We demonstrate our method on the four spectrographs of Las Cumbres Observatory's Network of Robotic \'Echelle Spectrographs (NRES), on the High Accuracy Radial Velocity Planet Searcher (HARPS) spectrograph, and on synthetic data. We obtain a velocity-equivalent precision of $\sim$\finalprecision~m/s on NRES. We achieve $\sim$1~m/s on HARPS, which agrees with the precision reported by the HARPS team. On synthetic data, we achieve the velocity precision set by Gaussian centroiding errors. Our algorithm likely holds for a wide range of spectrographs beyond the five presented here. We provide an open-source \textit{Python} package, \texttt{xwavecal}, which outputs wavelength calibrated spectra as well as the wavelengths of spectral features.
\end{abstract}

\received{October 17, 2019}
\accepted{May 12, 2020}

\maketitle

\section{Introduction}\label{sec:intro}
Detecting Earth-sized exoplanets at 1~AU around G-type stars requires relative radial-velocity (RV) precisions better than 10 cm/s \citep{Fischer_exoplanet_signals}. The \'echelle spectrograph is the current workhorse for such precision RV work, and new \'echelle spectrographs like ESPRESSO \citep{Faria+Adibekyan+2019+espresso, 2014AN....335....8P, espresso} and NEID \citep{NEID_2018AAS...23124608A} are rising to meet the 10 cm/s challenge. Wavelength calibration is the cornerstone of all precision RV work. Moreover, the number of exoplanet candidates is steeply increasing (for instance $10^4$ candidates are expected from the TESS primary mission, \citealt{2018arXiv180711129H}). Therefore, producing high quality RV measurements has several competing requirements for wavelength calibration: precision, automation, and propagation of uncertainty. Traditionally, \'echelle spectrographs have been wavelength calibrated with human aid at many steps. However, human involvement slows data throughput and traditional analysis techniques make formal propagation of error difficult. 

Wavelength calibration is defined by the wavelength solution: the mapping from position to wavelength. Finding the wavelength solution for an \'echelle spectrograph typically proceeds as follows. One locates the coordinates of spectral features on a wavelength calibration exposure, e.g. emission lines from a Thorium-Argon (ThAr) exposure or absorption features imprinted by an iodine cell. Then, with a sufficiently close initial guess, one refines the wavelength solution by minimizing the difference between identified spectral features and their matches in a reference wavelength list. Refining is straightforward if the initial guess is sufficiently good such that a majority of the features are matched to their true wavelengths in the reference list. This typically occurs when the initial guess is accurate to a few factors better than the typical line-spacing. The typical line spacing is $\sim1/2$ to $1$ \AA \, for the ThAr lamps used on NRES and HARPS.

Establishing an initial guess accurate to $\sim$10$^{-1}$ \AA \; is numerically challenging. A standard reference wavelength list has two or three features per Angstrom, and possesses tight clusters of lines where typical spacings are hundredths of an Angstrom. Thus an overall shift in the initial guess of just a few tenths of an Angstrom will cause \textit{most every} spectral feature to be identified with the \textit{wrong} line in the reference list, leading to a catastrophically incorrect final wavelength solution. The high density of lines in any calibration lamp spectrum will cause any goodness of fit metric to find thousands of local minima while traversing the wavelength solution's large parameter space. Accordingly, many algorithms require knowing the location and wavelengths of hundreds of lines a-priori, to sub-angstrom precision. The industry standard for establishing such a list of known positions and wavelengths is to match spectral features by-eye (e.g.~\citealt{CERES-2017PASP..129c4002B, CAFE2019arXiv190604060L}). Some use forward modelling and ray tracing of the optics \citep[e.g.~][]{HANLE2015ExA....39..423C} to find the list of approximate wavelengths. Many use IRAF (Image Reduction and Analysis Facility, \citealt{1986IRAF_standard_reference}) routines to find a list of known positions and wavelengths, and thereafter find the wavelength solution (e.g.~the Ond\v{r}ejov \'echelle spectrograph (\citealt{2016arXiv160204401G}) or WES (\citealt{WES2016PASP..128l5002G})). One such IRAF routine starts by identifying H$\alpha$ emission \citep{2016arXiv160204401G}. Iterating through the IRAF routines is time- and user-intensive, requiring up to tens of steps with user-feedback (see for instance, \citealt{2016arXiv160204401G}), and official support for IRAF ended in 2012. Propagating errors from IRAF numerical solutions is also difficult. Moreover, the lists of wavelengths and spectral features to include in the wavelength solution are often culled by trial and error, whereby combinations which seem to degrade the solution are removed. This trial-and-error process can take days and there is no guarantee that the best set of lines has been chosen. Moreover, spectra may have to be re-vetted later because features weaken relative to others as the calibration source ages \citep{2007MNRAS.378..221M}.

Some general tools for wavelength solutions have been created to reduce the human involvement required. The CERES pipeline \citep{CERES-2017PASP..129c4002B} has wavelength solutions for thirteen different instruments --- all based on a common set of routines. The new pipeline for the CAFE instrument \citep{CAFE2019arXiv190604060L} uses a modified version of the CERES wavelength solution. However, both the CERES and CAFE pipelines still require a-priori knowledge of the wavelengths of a set of spectral features and their pixel-positions on the detector. 

In this work, we present a novel algorithm that automatically produces an accurate and robust wavelength solution that does not require human intervention. With this, we can meet the demands that precision radial velocity measurements place on the wavelength solution.

This paper is structured as follows. We introduce cross-dispersed \'echelle spectrographs and their relevant properties in Section \ref{sec:intro_echelle}. In Section \ref{sec:methods}, we present our algorithm in detail. We stress the two step procedure integral to our method. We first solve for the scale-invariant portion of a low-degree wavelength solution via duplicated spectral features, then solve for the scale and higher-order components by matching the wavelengths of spectral features to their laboratory wavelengths. We show how to select a wavelength model as well. We discuss our open-source \textit{Python} implementation, \texttt{xwavecal} \citep{brandt_mirek_xwavecal_zenodo_2020}, in Section \ref{sec:discuss_code}. We show in Section \ref{sec:results} our algorithm's performance on synthetic data, and its performance on real data from both Las Cumbres Observatory's Network of Robotic \'Echelle Spectrographs (NRES) \citep{NRES-SPIE} and the High Accuracy Radial Velocity Planet Searcher (HARPS) spectrograph \citep{HARPS2003Msngr.114...20M}. We discuss considerations for adopting our algorithm in Section \ref{sec:discussion} and we conclude in Section \ref{sec:conclusion}.

\section{\'Echelle Spectrographs}\label{sec:intro_echelle}
A cross-dispersed \'echelle spectrograph has a high-dispersion optical element, usually a blazed grating, as its main dispersive element. The result is a large set of spatially long and thin diffraction orders that resemble stripes, which all lie on top of one another. The diffraction orders are separated by dispersing them along an axis perpendicular to the grating using a low-dispersion element such as a prism. The resulting spectrum resembles a ladder (une \'echelle) where each rung is a diffraction order ($50-100$ orders, or rungs, are visible on high resolution instruments). This spectrum is imaged by the spectrograph. Within each order the dispersion is high and the wavelength coverage (the free spectral range) is small. Because the free spectral range is small, the incident spectrum of the star, or calibration lamp, is not perfectly partitioned between the orders. Spectral features in the extremities of one diffraction order are often also present in one or more neighboring diffraction orders. The locations of these duplicated features encode the order to order variation in the wavelength solution. Our algorithm leverages the same features recorded repeatedly in consecutive orders.

Spectrographs are characterized by their resolving power $R = \lambda/ \Delta \lambda$. The majority of high-resolution ($R \gtrsim 50,000$) spectrographs image all the diffraction orders in at most two exposures: consider NRES, HARPS, ESPRESSO, The Hamilton \citep{Vogt_1987PASP_Hamilton}, ELODIE \citep{ELODIE-1996A&AS..119..373B}, or NEID. Some instruments, like the de-commissionied CRIRES \citep{Kaeufl+Ballester+etal_CRIRES_2004SPIE}, image only one diffraction order at a time. The optics are shifted to expose and record each diffraction order sequentially. Another example is NIRSPEC \citep{MartinNIRSPEC, Mclean1998DesignAD} in K-band, where capturing the full spectrum requires five exposures of its smaller-format infra-red chip. Whether or not every diffraction order is imaged simultaneously, there always exists a wavelength solution: the mapping between position in a diffraction order to wavelength $\lambda$.

Instruments that image all orders at once virtually guarantee, and all those in operation have shown, that the wavelength solution is a smooth function. By smooth we mean that the mapping from pixel to wavelength is free from high-frequency (e.g.~tens of pixels scale) variations. For instruments like CRIRES or NIRSPEC, which image the spectrum in many exposures, one could stitch the exposures together. The stitching will create high-frequency variations at the seams, but those could be calibrated out in principle. For such instruments, we assume those calibrations are possible and have been applied. We hereafter assume that we are calibrating a spectrum whose underlying wavelength solution is smooth, e.g.~spectra coming from NRES, HARPS, ESPRESSO, ELODIE, or NEID, or a mended CRIRES or NIRSPEC.

We showcase our algorithm on NRES and HARPS. NRES is a set of four spectrographs distributed between the northern and southern hemispheres. Each NRES spectrograph is a nearly identical $R \sim 50,000$ spectrograph that images the optical ($3800$ \AA \, to $8600$ \AA) on one 4k$\times$4k CCD. We show results from the NRES spectrograph at Sutherland, South Africa. HARPS has a resolving power of $\sim 120,000$ and images the optical across two CCDs. The HARPS detector is a mosiac of two 4k$\times$2k CCDs \citep{HARPSm0}. From the perspective of our algorithm, HARPS and NRES are similar. Their only major differences are that HARPS has twice the resolving power of NRES, and HARPS images the spectrum across two chips while NRES images the spectrum on one. We obtained HARPS calibration exposures from the ESO archive \footnote{\url{http://archive.eso.org/eso/eso_archive_main.html}} and NRES calibration exposures from the Las Cumbres Observatory archive \footnote{\url{https://archive.lco.global/}}.

We quantify the quality of a wavelength solution by the residuals between the inferred wavelengths of spectral features (from the wavelength solution) and their true, laboratory, wavelengths. An accurate wavelength solution has an unbiased distribution of residuals. The precision is often quantified by the scatter of those residuals (e.g.~the standard deviation if the residuals are Gaussian). We define the velocity-equivalent precision as the scatter converted to a velocity scatter $\sigma_v$ via the Doppler shift formula. Specifically, $\sigma_v^2$ is the variance of $c \Delta \lambda / \lambda$ where $\Delta \lambda$ is the residual between a spectral feature and its laboratory wavelength $\lambda$.

\section{Methods}\label{sec:methods}
We define our wavelength solution on an input spectrum that has been extracted and is therefore only a function of pixel coordinate $x$ and diffraction order index $i$. We define $\lambda(x,i)$ to be the wavelength solution: the mapping from pixel $x$ and order coordinate $i$ to wavelength $\lambda$. 
We index the diffraction orders from $i=0$ to $i=i_{\rm max}$ such that $0$ is the red-most order and $i_{\rm max}$ is the bluest, i.e.  $d \lambda /di < 0$. We assume that $d \lambda /dx > 0$ within an order, i.e., that wavelength increases from left to right. The model for our wavelength solution is based on the grating equation governing cross-dispersed spectrographs (e.g. \citealt{echelle_gratings_book}):
\begin{align}\label{eq:grating_equation}
    \lambda(\alpha, \beta) = \frac{d}{m_0 + i} (\sin\alpha + \sin\beta).
\end{align}
In Equation \ref{eq:grating_equation}, $d$ is the groove spacing and $m_0$ is the diffraction order number of the $i=0$ diffraction order, which we call the \textit{principle order number}. The principle order number is an integer, being a characteristic of the instrument (or of the instrumental setup, for configurable instruments). $\alpha$ and $\beta$ are the incident and outgoing angles for the diffracted light, measured with respect to the normal of the echelle grating. $\alpha$ and $\beta$ can be mapped into pixel coordinates. The smoothness assumption in Section \ref{sec:intro_echelle} is valid so long as the grating equation describes the instrument in question, because the grating equation itself is smooth. We transform from the angular coordinates and expand Equation \eqref{eq:grating_equation} in pixel and order index to obtain the following model.
\begin{align}\label{eq:wcs_full}
    \lambda(x,i) = \frac{1}{m_0 + i} \sum_{q=0}^{N_i} \sum_{c=0}^{N_x} a_{qc} P_q(i) P_c(x).
\end{align}
Equation \eqref{eq:wcs_full} is similar to the model used in the CERES pipeline \citep{CERES-2017PASP..129c4002B}. The $a_{qc}$ are free parameters with units of wavelength. $P_c(x)$ is the $c^{\rm th}$ degree basis function evaluated at $x$. We choose Legendre polynomials as our fit basis, though any orthogonal basis would be appropriate (Chebyshev polynomials for instance). $N_i$ and $N_x$ are the maximum degrees of the fit to the order-dependence and the pixel-dependence, respectively, of the wavelength solution. We use different $N_i, N_x$ during three stages of the wavelength calibration, gradually approaching $N_i = 5, N_x = 4$. $N_i = 5, N_x = 4$ works well for the spectrographs we have tested, but we discuss later how to find an appropriate $N_i, N_x$. Finding the wavelength solution amounts to solving for $a_{qc}$ and $m_0$.

\begin{figure} 
    \centering
    \includegraphics[width=\columnwidth]{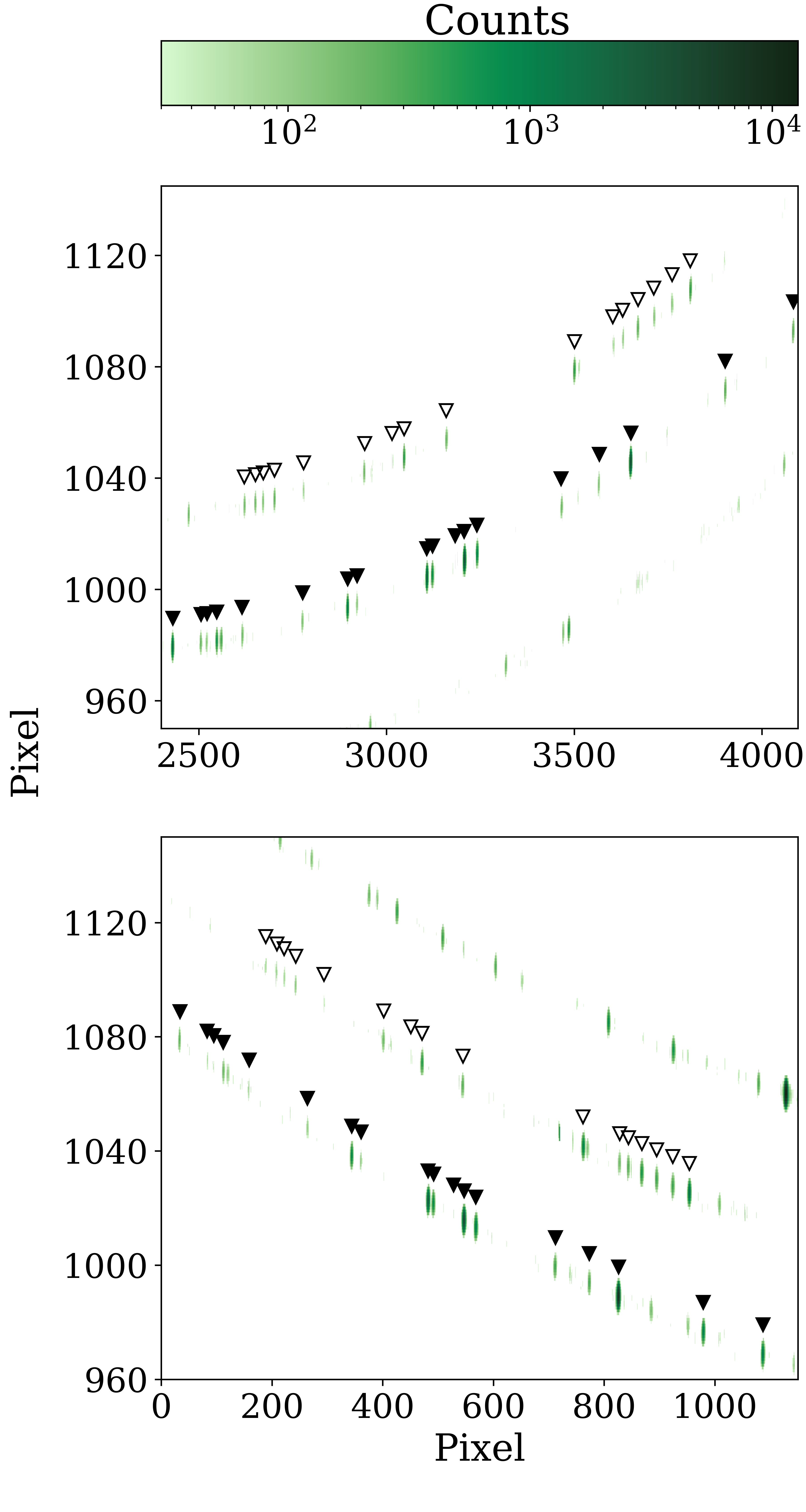}
    \caption{Reddest (top panel) and bluest (bottom panel) sections of three adjacent \'echelle orders. Each pair of duplicated emission lines are highlighted with hollow pointers for one diffraction order and filled pointers for the next order. The background has been removed for clarity.}
    \label{fig:overlapdetector}
\end{figure}

We solve for the free parameters of the wavelength solution by leveraging the fact that the wavelength coverage of consecutive diffraction orders of an \'echelle spectrograph overlaps. Many spectral features appear twice: in the blue side of the redder order and the red side of the bluer order. Figure \ref{fig:overlapdetector} displays this phenomenon with an un-extracted ThAr spectrum from NRES. Each of the $\sim$20 emission lines shown with pointers is present in two places: on the left edge of a redder order and on the right edge of the subsequent bluer order. High resolution instruments like NRES and HARPS have large regions of spectral overlaps and hundreds of duplicated features to use. For instance, about one third of all NRES spectral features have a duplicate at another location on the detector. The wavelengths of duplicated spectral features must be equal. By enforcing this consistency, we solve for all of the $a_{qc}$ in Equation \ref{eq:wcs_full} up to a single multiplicative pre-factor (for a given $m_0$).

Our approach works because the transformation that maps the pixel-position of a spectral feature to the position of its duplicate in another order is proportional to the ratio of the dispersion between the two orders (at those two locations). Forcing the wavelengths of two duplicated features to be equal constrains the wavelength solution independent of any reference wavelength list. This makes our initial estimate of the wavelength solution robust against features not tabulated in the reference list (e.g.~contamination from trace elements), artifacts from cosmic-ray hits, or bleed-over from strong features in adjacent orders.
 
The algorithm is structured as follows. We present our procedure for matching duplicated spectral features in Section \ref{sec:fitoverlap}. Using the overlaps, we solve for the wavelength solution up to a single multiplicative constant that we call the `global scale' (Section \ref{sec:unscaled_wcs}). We then introduce a reference list of laboratory wavelengths and find the global scale via a single parameter brute force search, which we describe in Section \ref{sec:scaled_wcs}. We refine the wavelength solution in Section \ref{sec:refine} by minimizing differences between each spectral feature and its closest match in the reference list. We treat $m_0$, the principle order number, as a known quantity throughout Sections \ref{sec:fitoverlap} -- \ref{sec:refine}. If $m_0$ is not known, i.e.~the instrument is new or uncharacterized, we iterate the whole procedure over a range of $m_0$ values, which we discuss in Section \ref{sec:m0}. We outline how to find an appropriate wavelength model in Section \ref{sec:choosing_model}. 

\subsection{The order overlaps}\label{sec:fitoverlap}
Hereafter, we assume the spectrum to be extracted; the spectrum is represented as a function of pixel $x$ and order $i$. We constrain the wavelength solution independent of any reference wavelength list by enforcing $\lambda(x,i) = \lambda(x', i+1)$ simultaneously for each spectral feature $(x, i)$ and its duplicate at $(x', i+1)$ in the subsequent order. Our goal then is to establish a large set of coordinates for the duplicated spectral features, i.e., many $[(x,i), (x', i+1)]$ where $\lambda(x,i) = \lambda(x', i+1)$.

\begin{figure}
    \includegraphics[width=\columnwidth]{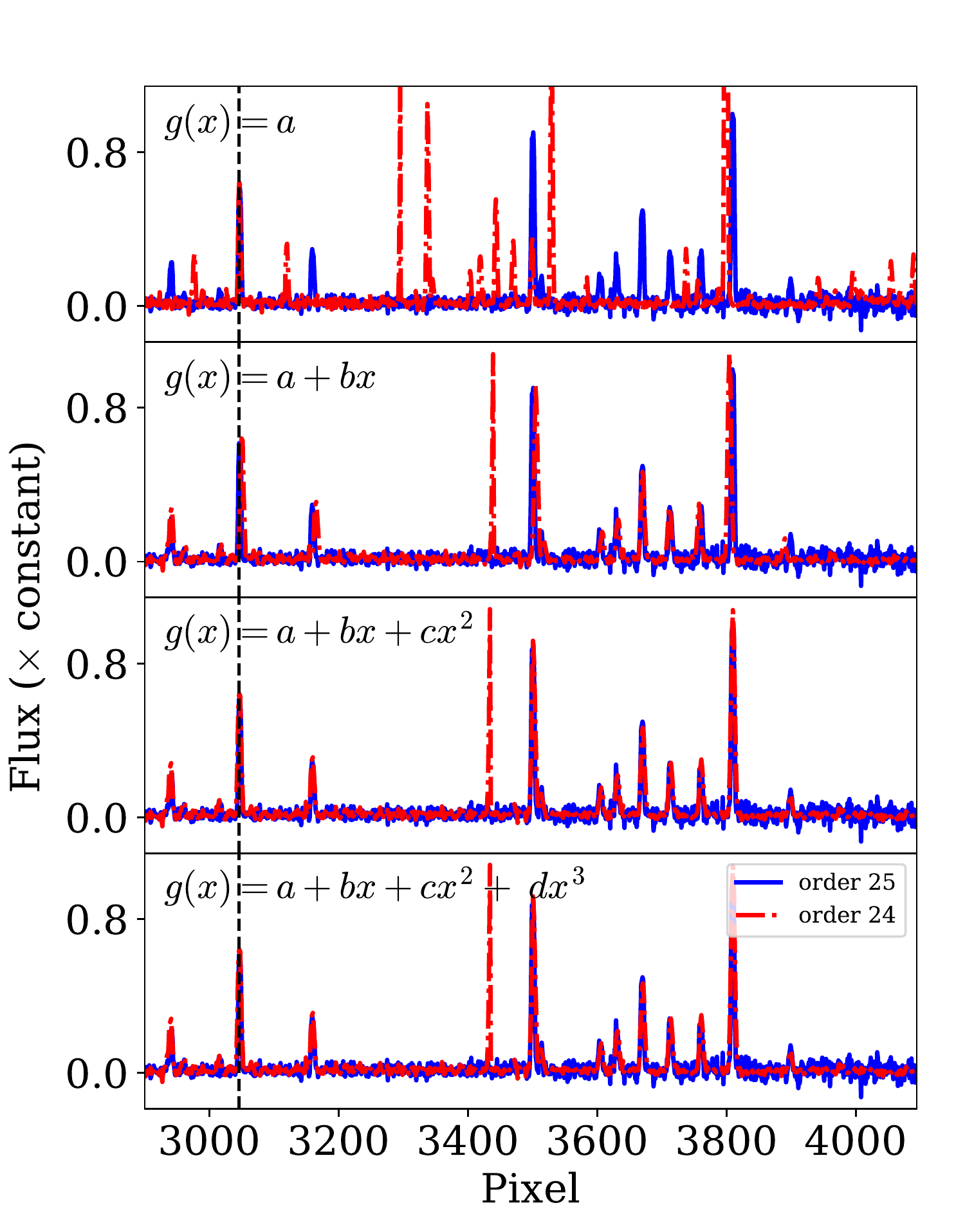}
    \caption{The overlap agreement between the red edge of the 25\textsuperscript{th} NRES diffraction order and the blue edge of the redder 24\textsuperscript{th} diffraction order. Both spectra have been blaze corrected (see Appendix A). The two orders are shifted in the top panel so that they are aligned at a duplicated feature at $x = 3050$, highlighted by the dashed black line. In each panel, the horizontal scale of order 24 has been transformed by a map $g(x)$ whose form is indicated in the upper left. The coefficients $a,b, \dots$ of $g(x)$ in each panel (except the uppermost) maximize the number of matched features.}
    \label{fig:overlapquality}
\end{figure}

To explain how to establish the set of matching coordinates, we examine an overlap closely. Figure \ref{fig:overlapquality} shows the extracted spectrum of two neighboring diffraction orders. The two orders are shifted in the top panel so that they are aligned at a duplicated feature at $x = 3050$. The two orders share many other spectral features but the dispersion $d \lambda / dx$ differs between them. The subsequent panels of Figure \ref{fig:overlapquality} show our corrections for the difference in $d \lambda / dx$ by distorting the horizontal scale with a polynomial mapping $g(x)$. A linear $g(x)$ improves agreement but is insufficient at the edges of the overlap. A quadratic mapping corrects for the difference in $d \lambda / dx$ as well as a cubic mapping. Extraneous features, such as the cosmic ray at $x \approx 3450$, do not pose a problem so long as $g(x)$ is free of small-scale noise, e.g., is represented as a quadratic or cubic polynomial. This is because any $g(x)$ that aligns such a bad feature with a feature on the blue side causes most every other blue feature to become misaligned and therefore produces fewer matched features than the best-fit solution. Moreover, $g(x)$ must be free of small pixel scale variations because the wavelength solution is as well. The $g(x)$ encode $d\lambda/dx$; they vary smoothly from order to order. We use $g_i(x)$ to designate the overlap of order $i$ with order $i+1$.

We define \textit{fitting an overlap} as finding the coordinate mapping $g_i(x)$ that corrects for $d \lambda / dx$ between two overlapping orders, e.g.~finding the coefficients $a_i, b_i, c_i$ of $g_i(x)$. A successful fit causes every duplicated spectral feature to align within a fraction of a pixel. Given many $g_i(x)$, we calculate and use the pixel-coordinates of all duplicated features to solve for the wavelength solution up to an unknown prefactor (the global scale).
\par
Within the overlap region of two orders $i$ and $i+1$, we must have a $g_i(x)$ such that
\begin{align}\label{eq:overlap_condition}
    \lambda(x, i+1) = \lambda(g_i(x), i). 
\end{align}
In practice, $x$ is the coordinate of a spectral feature in the $i+1$ order and $g_i(x)$ would be the coordinate of a spectral feature in order $i$. Equation \eqref{eq:overlap_condition} above applies for all points $x$ in the set of overlapping coordinates.

\par

We assume a pair of one-dimensional spectra from adjacent orders, $i$ and $i+1$, have spectral features with flux weighted centers at $x_n$ and $x_m$, respectively, with $n=0,1,..., m = 0,1,...$. We define a spectral feature from order $i$ as matched if its flux-weighted center $x_n$ is within one pixel of the flux-weighted center of the duplicate under transform $g(x_m)$, i.e. $x_n = g(x_m) \pm 1$. We find the best $g_i(x)$ by setting $g_i(x) = a_i + b_i x + c_i x^2$ and optimize $a_i, b_i$ and $c_i$ to maximize the number of matching spectral features between the two orders. We describe this algorithm in detail in Appendix A. Finding a good $g_i(x)$ for as many overlaps as possible concludes fitting the overlaps. 

Given the $g_i(x)$ for an overlap between order $i$ and $i+1$, $[(g_i(x), i), (x, i+1)]$ is a coordinate pair that maps to a common wavelength (i.e.~that satisfies Equation \eqref{eq:overlap_condition}). Each coordinate pair is one data point which constrains our wavelength solution. Across every overlap, we evaluate the coordinate pairs $(g_i(x), i), (x, i+1)$ for all duplicated spectral features, yielding hundreds of coordinate pairs that we use to constrain the entire wavelength solution save one remaining parameter: the global scale.

\subsection{Fitting the unscaled wavelength solution}\label{sec:unscaled_wcs}
In this Section, we solve for the coefficients of the wavelength solution (Equation \eqref{eq:wcs_full}) using spectral features only from orders whose overlaps were successfully fit. To fit Equation \eqref{eq:wcs_full}, one must pick a maximum degree for the order and pixel dependence of $\lambda(x,i)$. By default in \texttt{xwavecal}, $N_x = 2$ and $N_i = 2$ because those sufficed for reproducing the overlaps over a range of twenty orders. Too large of either $N_x$ or $N_i$ at this stage can lead to over-fitting. To constrain our solution, we define a \textit{global scale} $K$ such that $a_{qc} = K b_{qc}$, where $b_{qc}$ are unitless coefficients. Comparing Equation \eqref{eq:factored_wcs} with the grating equation, Equation \eqref{eq:grating_equation}, reveals that $K$ should nearly equal twice the groove spacing $d$ if the spectrograph is in the Littrow configuration \citep{echelle_gratings_book}. For fixed $m_0$, the $b_{qc}$ describe the scale-independent shape of the wavelength solution. The wavelength solution that we constrain using the overlaps is
\begin{align}\label{eq:factored_wcs}
    \lambda(x,i) = \frac{K}{m_0 + i} \left( 1  + \sum_{q=0}^{2} \sum_{c=1}^{2} b_{qc} P_q(i) P_c(x) \right).
\end{align}
Note that we have factored out the global scale constant $K$. The overlaps constrain $d\lambda / dx$ and therefore do not constrain any $b_{q0}$ ($x$-independent) coefficients. We thus force $b_{00} = 1$ and $b_{10} = b_{20} = 0$ so that we find non-trivial solutions for the free $b_{qc}$. This assumption is equivalent to assuming that the mean wavelength of a diffraction order evolves like $1/(m_0 + i)$, i.e. that the grating equation well approximates the wavelength solution. We relax this constraint later on.

From fitting the overlaps in Section \ref{sec:fitoverlap}, we have a set of pixels from neighboring orders that must map to the same wavelength. In other words, we have the following set (denoting $g_i(x) = x'$ for the $i, i+1$ overlap.)
\begin{align}
    \mathcal{O} = \{ ((x,i), (x', i+1)) \text{ s.t. } \lambda(x,i) = \lambda(x', i+1) \}.
\end{align}
The number of overlaps fit is the number of well-constrained coordinate mappings from the blue edge of a redder diffraction order to the red edge of the neighboring bluer order. If thirty overlaps were fit, each with ten matched features, then $\mathcal{O}$ would have 300 elements. Thus, Equation \eqref{eq:factored_wcs} is over-determined (if more than $\sim$5 overlaps were fit) and we use least squares to solve for the best fit coefficients of $\lambda(x,i)$ up to a multiplicative prefactor. Because we fit each overlap with a quadratic, the problem is constrained by many fewer parameters than the 300 elements in $\mathcal{O}$. We retain $\mathcal{O}$ when we fit, which weights orders by the number of features that match within the overlap. Equating $\lambda(x,i)$ to $\lambda(x', i+1)$ we find the following $|\mathcal{O}|$ equations for the $b_{qc}$:
\begin{align}\label{eq:solve_wcs_for_overlap}
     & \sum_{q=0}^{2} \sum_{c=1}^{2} \left( \frac{P_q(i+1)  P_c(x')}{m_0 + i + 1} - \frac{P_q(i)  P_c(x)}{m_0 + i} \right) b_{qc} \\ \nonumber
     &=\frac{1}{m_0 + i} - \frac{1}{m_0 + i + 1} \\ \nonumber
     & \hspace{70pt}  \forall ((x,i), (x', i+1)) \in \mathcal{O}
\end{align}
where we divided out the global scale $K$ which was common to both sides. Equation \eqref{eq:solve_wcs_for_overlap} is an over-determined system linear in the six $b_{qc}$. We solve for the best-fit $b_{qc}$ with least squares through the \textit{Python} package \texttt{numpy.linalg.lstsq} \citep{numpy1, numpy2}. Assuming a known $m_0$, we now have all the parameters of a low-order wavelength solution except for the global scale factor $K$.

\subsection{Finding the scale of the wavelength solution}\label{sec:scaled_wcs}
To find the global scale $K$, we introduce a reference wavelength list and scale the unscaled wavelength solution to match the structure of the list. Given the wavelength solution up to a global scale $K$, any goodness of fit metric for $K$ will be littered with local minima. Minimization routines that rely on gradients like Nelder-Mead or Levenberg-Marquardt will cling to these local minima. This problem is thus intractable by conventional minimization routines unless they are supplied an extremely accurate initial guess such that the nearest local minimum is at the correct $K$ value and therefore optimization is convex. Otherwise, the minimization routine must have a way to climb out of local minima, e.g. simulated annealing or basin hopping. We find a sufficiently accurate starting guess for $K$ with a brute force search. We search over a grid spaced such that the brute force result will be close enough to the correct $K$ for subsequent optimization to be convex. We then refine $K$ with Nelder-Mead as implemented in \texttt{scipy.optimize.minimize} \citep{scipy}.

We need three quantities to find $K$: a grid spacing that guarantees that we estimate $K$ accurately enough; a range of values in $K$ to search over; and a metric that will be minimized by the correct value of $K$. Our metric, $\metric(K)$, is sum of the squared residuals between laboratory wavelengths and each spectral feature's wavelength under the trial model given by $K$. We define the metric in detail in Appendix B.

We calculate the requisite grid spacing using the fact that the final precision on $K$ must yield a wavelength solution with a precision better than the average line spacing. We illustrate this calculation with an $R = 50,000$ spectrograph and typical values for a ThAr emission line list. We always have that $\delta \lambda / \lambda \sim \delta K / K$ since $\lambda \sim K/m_0$. If the average spacing in the line list is $\delta \lambda \sim 1$ \AA \, at 5000 \AA, then $\delta K / K \lesssim 1/5000 \sim 2 \cdot 10^{-4}$. If the scale estimate from Equation \eqref{eq:order_of_magnitude_K} is ${\sim}4 \cdot 10^5$ \AA, we should search for the global scale in steps no larger than $\delta K  \lesssim 10^{-4} \cdot 8 \cdot 10^5 \text{\AA} \, \approx 100$ \AA \, to achieve a precision better than 1 \AA. The \texttt{xwavecal} default is $\delta K = 10$ \AA \,, which achieves a precision of $0.1$ \AA \, on $R = 50,000$ optical spectrographs. Additionally, the $\delta K$ estimate also tells us that over a range of $5 \cdot 10^4$ \AA \, in $K$, there are on the order of $5 \cdot 10^4/ \delta K \approx 500$ local (erroneous) minima. We have confirmed that estimate by computing $\metric(K)$ in practice (See Figure \ref{fig:chisqvslambdascale}).

We find the search range over $K$ by first estimating $K$ as follows. If the spectrograph spans $i_{\rm max}$ unique diffraction orders and the wavelength span across all orders is $\Delta \lambda$, then inverting the grating equation gives a global scale of roughly 
\begin{align}\label{eq:order_of_magnitude_K}
    K \sim \Delta \lambda \cdot (1/m_0 - 1/(m_0 + i_{\rm max}))^{-1} \text{ \AA}.
\end{align}
The estimate from Equation \eqref{eq:order_of_magnitude_K} is typically accurate to within $5$\% and serves as a central point for the brute force search. Alternatively, $K$ will also be nearly equal to twice the groove spacing if the spectrograph is in Littrow configuration \citep{echelle_gratings_book}, and so the groove spacing from the grating manufacturer is a suitable estimate as well. \texttt{xwavecal} by default searches above and below the estimate by a factor of two.

Using the aforementioned range of $K$ values, grid spacing, and metric $\metric(K)$, we brute-force search for the $K$ value which minimizes $\metric(K)$. See Appendix B for details. We then refine the $K$ estimate using Nelder-Mead as implemented in \texttt{scipy.optimize.minimize}. With the global scale $K$ in hand, we multiply our dimensionless $b_{qc}$ by $K$ and recover the dimensionfull $a_{qc}$. These $a_{qc}$ are constrained over a limited range of orders. We next constrain the wavelength solution using every spectral feature from every order, refining all of the parameters in the wavelength solution.

\subsection{Refining the wavelength solution}\label{sec:refine}
Before we refine the wavelength solution, we remove the constraints introduced in fitting the overlaps: we no longer force $a_{00}$ to be unity nor the other $a_{q,0}$ to be zero. Therefore we have $(N_x + 1)(N_i + 1)$ degrees of freedom. We formally require only $(N_i + 1) \times (N_x + 1)$ spectral features to constrain the parameters of the wavelength solution (the $a_{qc}$ of Equation \eqref{eq:wcs_full}). Many more features yield a more reliable solution due to measurement error on each individual feature. However, as long as we have enough features ($\sim6$ per overlap or $\sim20$ per order) to constrain the overlaps, then there are more than enough constraints to refine the wavelength solution.

We refine the wavelength solution in two steps: 
\begin{itemize}
    \item [1.] We add sets of features order-by-order until the wavelength solution is constrained over all of the orders.
    \item [2.] We add degrees of freedom one-at-a-time to the wavelength solution until it reaches the desired complexity.
\end{itemize}
Step 1 slowly constrains the wavelength solution which helps it converge to the correct solution. Step 2 gradually makes the wavelength solution more complex, which mitigates early over-fitting. This two step procedure is computationally cheap. It requires roughly one hundred solving iterations in total.

In both steps, we solve for the $a_{qc}$ coefficients with Iteratively Re-weighted Least Squares (IRWLS) minimization \citep{lawsons, lawsons_revisit}. For each spectral feature (indexed by $n$) at coordinate $(x_n, i_n)$ we find the reference wavelength closest to $\lambda(x_n, i_n)$, which we call $\lambda_{{\rm ref}, n}$. We enforce $\lambda(x_n, i_n) = \lambda_{{\rm ref}, n}$ for all $X$ spectral features and solve the $X$ equations, which are linear in the $a_{qc}$. We use binary weights $w_n$ in IRWLS and assign $w_n=0$ to outliers. We define outliers at each iteration using a settable\footnote{See Table \ref{table:xwavecal_parameters}, or the end of this sub section, for recommended values.} threshold on the median absolute deviation (m.a.d.) rather than the standard deviation often used in sigma clipping. For Gaussian data, ${\rm m.a.d.} \approx 0.67\sigma$, but the m.a.d.~is resistant to outliers which normally would degrade the solution.

In step 1, we initialize the weights $w_n$ by ignoring features $\lambda(x_n, i_n)$ outside the range of orders successfully fit in Section \ref{sec:fitoverlap} (we set their $w_n=0$). We only consider features $(x_n, i_n)$ with error $| \lambda(x_n, i_n) - \lambda_{{\rm ref}, n} | $ less than $\kappa$ m.a.d.'s from zero. We set outlier weights $w_n=0$ for one iteration only and recompute outliers after every iteration. We then add features from adjacent orders, one order from both the blue and red at a time, and solve again for the $a_{qc}$ coefficients. If our solution was constrained only between orders 10 and 20, our next iteration would refine the solution with features (except outliers) from all orders between 9 and 21. We iterate until every identified feature from every order constrains the wavelength solution. For NRES and HARPS, the wavelength solution after this stage yields residuals with the m.a.d. less than $10^{-1}$ \AA. This concludes finding the initial guess to the wavelength solution.
\par
In final refinement (step 2), we gradually inject more polynomial degrees of freedom into Equation \eqref{eq:wcs_full}. Slowly adding degrees of freedom prevents over-fitting and helps convergence. We ignore $\kappa'$-m.a.d. outliers instead of $\kappa$-m.a.d. outliers. We solve for the $a_{qc}$ with IRWLS as before. After converging, we add one new degree of freedom (one new $a_{qc}$). For example, after converging with $N_i = 2$ and $N_x = 2$, we add $a_{30} P_3(i) P_0(x)$ to Equation \eqref{eq:wcs_full}. We solve until we converge and then iterate; adding one more degree of freedom after each convergence until we reach Equation \eqref{eq:wcs_full} with the user-set final model. The \texttt{xwavecal} default final model has $N_i = 5$ and $N_x = 4$. The clipping parameters $\kappa$ and $\kappa'$ can be set the user, but we recommend the default values of six and four, respectively. This concludes the wavelength solution.

\subsection{The principle order number $m_0$}\label{sec:m0}

The preceding sections assumed a value for the principle order number $m_0$. The principle order number should be known. If the principle order number is not known, we can iterate the entire procedure for a range of $m_0$. The correct $m_0$ will minimize the median absolute deviation after final refinement (in principle, the median absolute deviation can be replaced by any outlier-resistant measure of the scatter).
\par
\begin{figure}[]
    \includegraphics[width=\columnwidth]{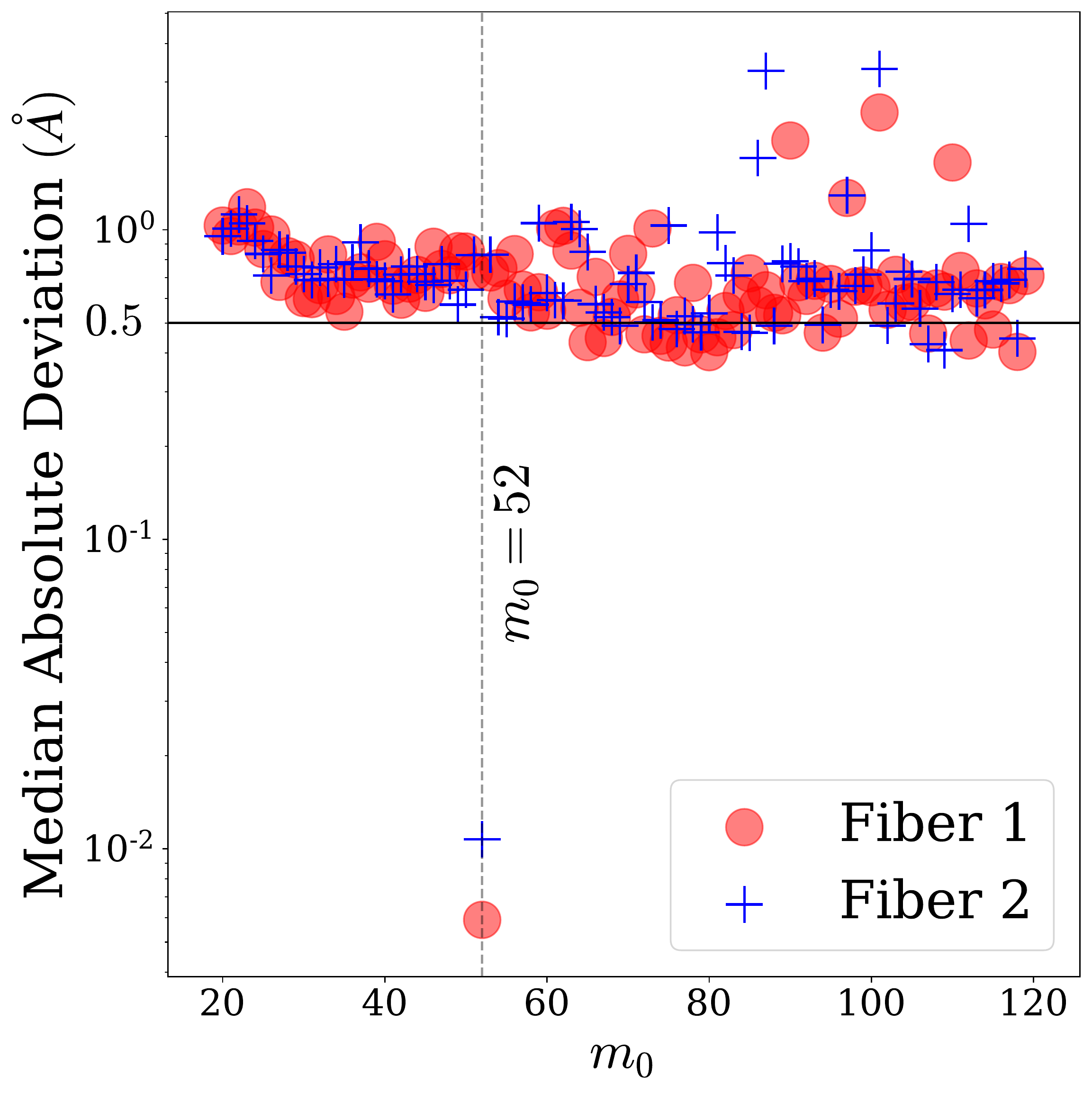}
    \caption{The median absolute deviation of the residuals from the final wavelength solution as a function of assumed principle order number $m_0$. The true principle order number is 52 and is the global minimum. We show the independent solutions for both fibers in this NRES calibration exposure.}
    \label{fig:m0_merit}
\end{figure}

 
Our search is constrained because $m_0$ is an integer and by design $m_0 \lesssim 200$ so that the instrument is efficient. For example, $m_0 {\sim}24$ for FEROS \citep{FEROSm0}, 88 for ELODIE \citep{ELODIE-1996A&AS..119..373B}, 52 for NRES, and 89 for HARPS \citep{HARPSm0, HARPS2003Msngr.114...20M}. We brute force search for $m_0$ between 5 and 200, using the final scatter of the wavelength solution as our metric. The parameters of the wavelength solution, e.g. $K$ and the $a_{qc}$, will depend on the trial value of $m_0$ and so one must run the entire wavelength solution algorithm (including the $K$ brute force search) for each value of $m_0$. Although this brute force search over 200 $m_0$ values is expensive (taking roughly 30 minutes using a single CPU-core), we only need to do it once per spectrograph because the principle order is intrinsic to the instrument design. Figure \ref{fig:m0_merit} shows the median absolute deviation of the residuals after final refinement for both fibers on the same calibration exposure as a function of $m_0$. The true principle order number for NRES, $m_0 = 52$, yields a scatter that is a factor of 100 smaller than the next best fit. 

\subsection{The final wavelength model}\label{sec:choosing_model}
There are a variety of wavelength models used in the literature, even for instruments of similar resolving power. For example: the ELODIE spectrograph pipeline \citep{ELODIE-1996A&AS..119..373B} used $N_i = 5$ and $N_x = 3$; the EXPRES pipeline \citep{Petersburg+etal_EXPRES_2020} has $N_i = 6$ and $N_x = 6$; CERES chooses $N_i$ and $N_x$ by visually inspecting if there is any structure in the residuals of the wavelength solutions \citep{CERES-2017PASP..129c4002B}; and \cite{Donati+Semel+Carter+etal_1997} used $N_i = 12$ and $N_x = 3$.

\begin{figure}
    \centering
    \includegraphics[width=\columnwidth]{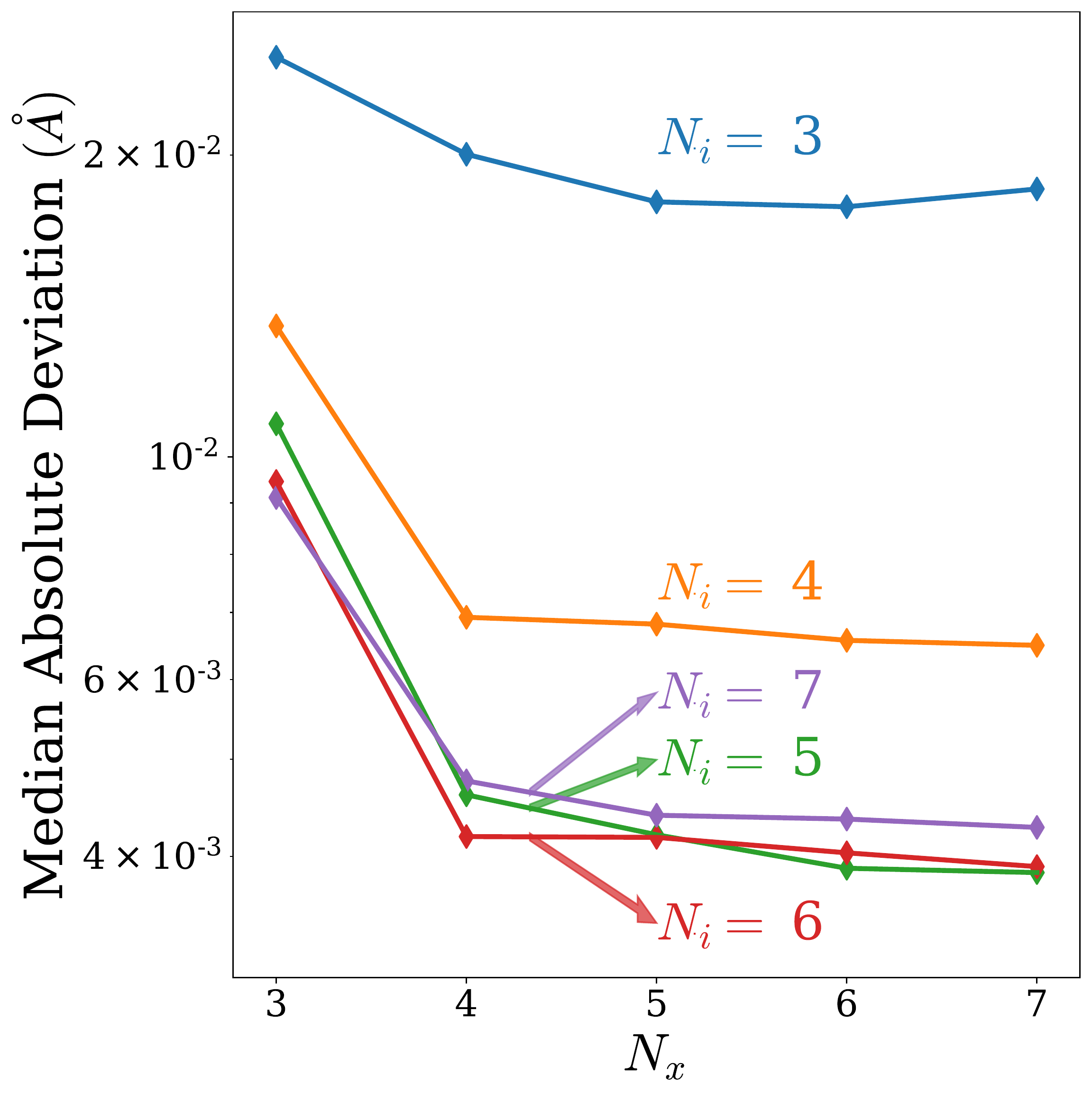}
    \caption{The median absolute deviation of the residuals between the wavelengths of spectral features and their reference values, as a function of the polynomial ($N_x$) and order ($N_i$) degrees of freedom in the final wavelength model. We have labelled curves of constant $N_i$ and distinguished closely lying curves with arrows when necessary.}
    \label{fig:wavelength_models}
\end{figure}

We recommend that the user selects the least complex model that reaches a set precision threshold on the residuals. For example, Figure \ref{fig:wavelength_models} illustrates the residuals from different models on NRES. Models higher order than $N_i = 5$ and $N_x = 4$ do not yield reduced scatter: one gains a factor of five in precision on NRES by adding ten degrees of freedom to move from a model with $N_i = 3, N_x = 4$ to one with $N_i = 5$ and $N_x = 4$. One does not gain any precision by adding another ten degrees of freedom to move $N_i = 5$ to $N_i = 7$. This suggests that the data do not meaningfully constrain those higher order terms and that they should therefore be excluded from the model. Figure \ref{fig:wavelength_models} shows that $N_i = 5$ and $N_x = 4$ is the least complex, highest precision model of those tested.

{}

\section{ \texorpdfstring{\MakeLowercase{\texttt{xwavecal}}}{xwavecal}}\label{sec:discuss_code}
\texttt{xwavecal} \citep{brandt_mirek_xwavecal_zenodo_2020} is our open source\footnote{\url{https://github.com/gmbrandt/xwavecal}} \textit{Python} implementation of our algorithm, which we built on the Template Method pattern \citep{design_patterns} and the framework of \texttt{BANZAI}. \footnote{\url{https://github.com/lcogt/banzai}} \citep{banzai_spie, curtis_mccully_banzai_zenodo_2018}. One can install \texttt{xwavecal} with \texttt{pip install xwavecal}. \texttt{xwavecal} can wavelength calibrate in two ways:
\begin{itemize}
    \item [1.] The command line entry-point \texttt{xwavecal\_reduce}. This requires a \texttt{.ini} configuration file with the instrument-specific configurations  and a properly formatted, extracted, calibration spectrum (e.g.~from ThAr). It saves the wavelength calibrated input spectrum, where each pixel has a wavelength assigned.
    \item [2.] The \texttt{wavelength\_calibrate} function. This function takes two sets of data and a set of arguments. The data are the measured feature positions (pixel, order, flux) and a reference wavelength list. The arguments are the same data that are stored in the configuration files. \texttt{wavelength\_calibrate} returns the wavelengths of the input features.
\end{itemize}
Method 1 is closer to a full-reduction pipeline and would be suitable for users who want to wavelength calibrate a large set of spectra. Method 2 is suitable for users who are interested in incorporating the central routines of \texttt{xwavecal} into a pipeline of their own. For instance, one could use method 2 as a fallback in a pipeline to generate a fresh wavelength calibration if they detect an instrument shift. Both methods require instrument-specific information. For method 1, the user will populate a configuration \texttt{.ini} file for their instrument. For method 2, the user would supply that information to the \textit{Python} function as keyword arguments.

To run \texttt{xwavecal} via method 1, one needs to input a \texttt{.fits} file containing an extracted calibration spectrum. An extracted spectrum means a table of 1D spectra, where each row is the 1D spectrum of a diffraction order. The spectrum needs 1-sigma flux errors for each point. The format of the input spectra is described in more detail in the documentation. To marginally improve performance, one can provide a blaze corrected version of the same spectrum (see Appendix A for a description of blaze correction).

After providing the input spectrum and modifying the configuration file, method 1 of \texttt{xwavecal} outputs: a wavelength calibrated spectrum; a table of spectral feature wavelengths, pixel, and order positions and errors; and a table of overlap information. The overlap table consists of each matched spectral feature and the pixel position of its duplicate in the neighboring diffraction order. The \textit{ESO ThAr atlas} is included with \texttt{xwavecal} and will suffice for most high resolution optical instruments.
\par
Possibly the most useful output of both methods of \texttt{xwavecal} is the table of spectral feature positions and reference wavelengths, which is what most pipelines require (e.g.~CERES) for their own wavelength solution. \texttt{xwavecal} can quickly and automatically obtain accurate wavelengths for every line on the detector that \textit{any} reduction pipeline can use. With \texttt{xwavecal} and a single-core of a 3.4 GHz laptop CPU and 8 gigabytes of RAM, the entire wavelength solution takes $\lesssim$30 seconds to calibrate one spectrum with 67 diffraction orders and $\sim$1500 spectral features. The most time-intensive step is fitting the overlaps.

We encourage additions to \texttt{xwavecal}. The code is modular: each reduction step is a stage which can be easily disabled from the configuration file. A new stage can be added with a single line of text in the configuration file, without modifying the \texttt{xwavecal} source code in any way.

Most every parameter discussed in this work can be changed, but very few need to be changed. Table \ref{table:xwavecal_parameters} lists the \texttt{xwavecal} parameters required in any configuration file. The table includes a short description, the default value for each parameter, whether or not one should expect to change the parameter for an $R=20,000 - 100,000$ spectrograph, and the expected consequence of optimizing that parameter. We have omitted parameters which: pertain to reading and writing files for \texttt{xwavecal} method 1 (e.g. the reference list file path); or do not need to be changed for a new instrument. We refer the inclined reader to the documentation with the source code for a thorough discussion of every parameter.

There are only six parameters that we recommend changing in Table \ref{table:xwavecal_parameters}. Optimizing five of these parameters serve only to reduce runtime. Moreover, these six can be obtained from an inspection of a calibration frame of that instrument, or from the instrument's online information page.

\begin{table*}
    \centering
    \begin{threeparttable}
    \caption{Descriptions of important \texttt{xwavecal} parameters.\label{table:xwavecal_parameters}}
    \begin{tabular}{LlCLLL}
        \toprule \\
        Name in config & Name & Description & Default$^{1}$ & Change? &  Result of Optimizing\\
        \hline \\
$^{2}$overlap\_min\_peak\_snr     & \nodata    & Features with a signal to noise above this threshold will be used in the overlap algorithm.          & 5                         & No              & Performance                     \\ \hline
$^{\text{A}}$flux\_tol                          & \nodata    & Flux agreement threshold for two features to be accepted as a match.                                   & 0.2                       & Maybe$^{\star}$ & Speedup \& Performance                         \\ \hline
$^{\text{A}}$min\_num\_matches          & $F$    & Minimum number of matching features to accept an overlap fit as valid.                    & 6                         & No              & \nodata                          \\ \hline
max\_red\_overlap                  & \nodata    & Maximum size in pixels of the overlap on the red side of each order. Half the detector is good.      & 1000                      & Yes             & Speedup                         \\ \hline
max\_blue\_overlap                 & \nodata    & Maximum size in pixels of the overlap on the blue side of each order. Half the detector is good.     & HARPS:~1000 NRES:~2000 & Yes             & Speedup                         \\ \hline
min\_num\_overlaps                 & \nodata    & Minimum number of successfully fit overlaps in order to proceed with calibration.                    & 5                         & No              & \nodata                          \\ \hline
global\_scale\_range               & \nodata   & The multiplicative search range above and below the initial guess for the global scale.              & (0.5, 2)                  & No              & Speedup                         \\ \hline
global\_scale\_spacing               & $\delta K$   & The grid spacing in Angstroms for the global scale search              & 10                  & No              & Speedup                         \\ \hline
approx\_detector\_range\_angstroms & \,    & Approximate spectral range of the spectrograph in angstroms.                                          & See Note 3.               & Yes             & Better initial guess for $K$. \\ \hline
approx\_num\_orders                & \nodata    & Approximate number of diffraction orders in the input spectrum.                                      & HARPS:~45 NRES:~67      & Yes             & Better initial guess for $K$. \\ \hline
m0\_range                          & \nodata    & Range of $m_0$ to search over, if $m_0$ is unknown.                                                  & (10, 200)                 & No              & Speedup                         \\ \hline
principle\_order\_number           & $m_0$     & The true diffraction order index of the first diffraction order in the spectrum.                     & NRES:~52     & Only if known.  & \nodata                          \\ \hline
initial\_mad\_clip          & $\kappa$  & Outlier clipping parameter for the initial refinement stage.                                         & 6                         & No              & Performance                     \\ \hline
final\_mad\_clip            & $\kappa'$ & Outlier clipping parameter for the final refinement stage.                                           & 4                         & No              & Performance                     \\ \hline
$^{2}$min\_peak\_snr              & \nodata    & Features with a signal to noise above this threshold will be used to refine the wavelength solution. & 20                        & No              & Performance                     \\ \hline
initial\_wavelength\_model         & \nodata    & The wavelength model that the overlaps will constrain.                                               & See text.                 & No              & Performance                     \\ \hline
intermediate\_wavelength\_model  & \,    & The wavelength model that will be constrained by slowly adding features from the entire spectrum.     & $N_i=2, N_x=2$                 & No              & Performance                     \\ \hline
final\_wavelength\_model           & \nodata    & The wavelength model to constrain using every feature in the spectrum.                                & $N_i=5, N_x=4$                 & No              & Performance                     \\ \hline
        \end{tabular}{}
        \begin{tablenotes}
        \small
            \note{~`Speedup' means reduced algorithm execution time without improved performance. `Performance' implies a reduced chance of the wavelength solution failing.}
            \item[] $^{1}$ The parameters used in this work to calibrate NRES and HARPS. These are not necessarily optimal.
            \item[] $^{2}$ These parameters do not apply if you use method 2 and provide \texttt{xwavecal} with a list of feature positions. Note that these parameters can be set equal, but it marginally helps performance to have many low signal-to-noise features in the overlap fitting algorithm.
            \item[] $^{3}$ 1800 for a single chip of HARPS. 6000 for NRES.
            \item[] $^{\star}$ We recommend 0.2 if the feature fluxes are blaze corrected, and 0.6 if they are not.
            \item[] $^{\text{A}}$ See appendix A for where this parameter is used.
        \end{tablenotes}
    \end{threeparttable}
\end{table*}

\newpage

\section{Results}\label{sec:results}
We calibrate real data from NRES and HARPS with \texttt{xwavecal} in Section \ref{sec:residuals}. We then test our algorithm on synthetic data (Section \ref{sec:synthetic_data}) with contamination and centroiding errors. We determine the best-case precision of our algorithm, its contribution to the error budget, and the conditions under which it can fail.

\subsection{NRES and HARPS}\label{sec:residuals}

We calibrated NRES and HARPS using the default parameters of \texttt{xwavecal} and the \textit{ESO ThAr atlas}\footnote{\url{http://www.eso.org/sci/facilities/paranal/instruments/uves/tools/tharatlas/thar_uves.dat}} for our reference wavelength list. We showcase only the NRES unit at Sutherland, South Africa, but the results are nearly identical for the other three NRES units. For HARPS, we calibrated one CCD which happened to contain the blue-most ${\sim} 45$ orders. We did not know a-priori whether the CCD we chose was the red or blue half of the full detector. Both instruments' calibration spectra have an average signal to noise per feature of roughly 50, and were calibrated using the default \texttt{xwavecal} configuration shown in Table \ref{table:xwavecal_parameters}. 

\begin{figure*}
    \includegraphics[width=0.5\textwidth]{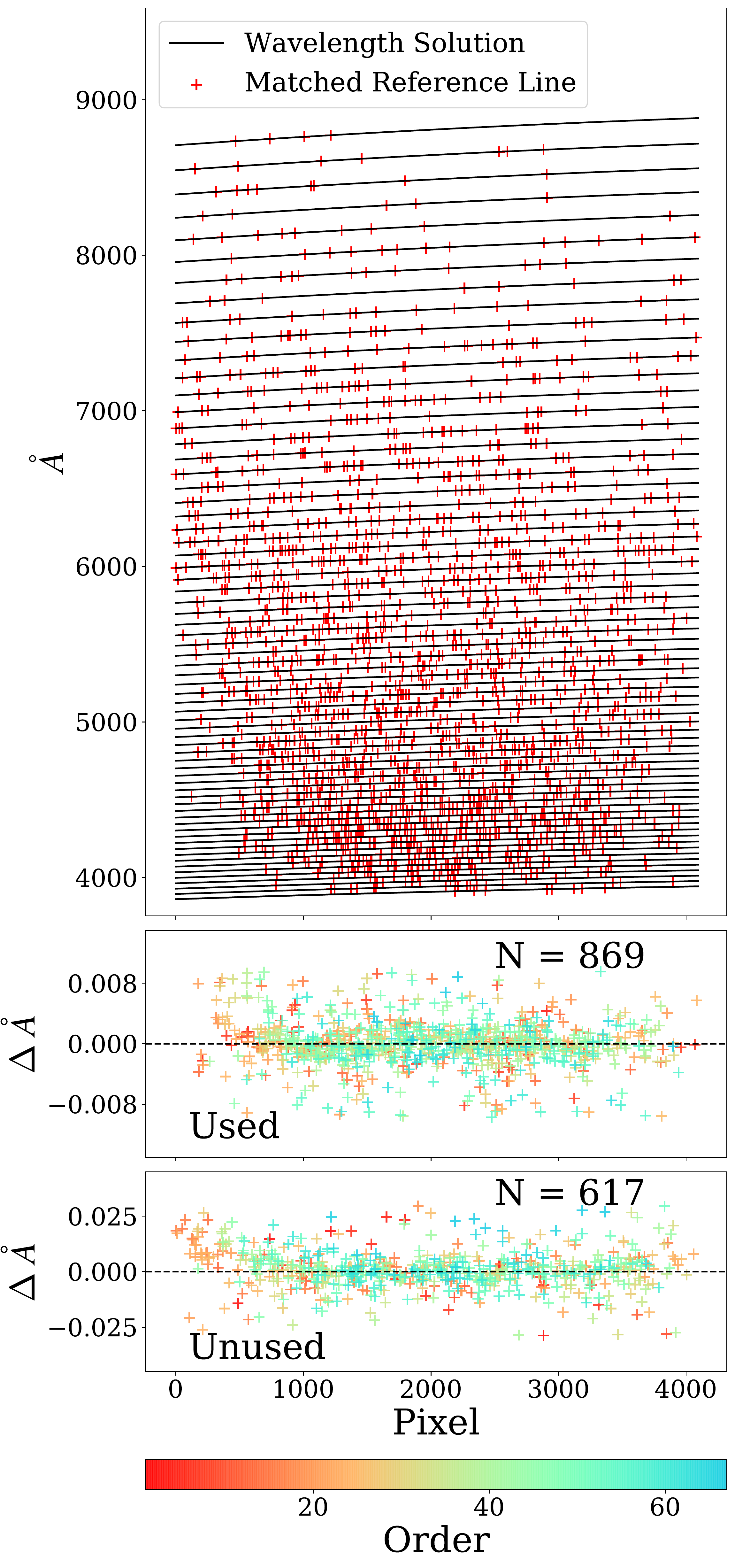}        \includegraphics[width=0.5\textwidth]{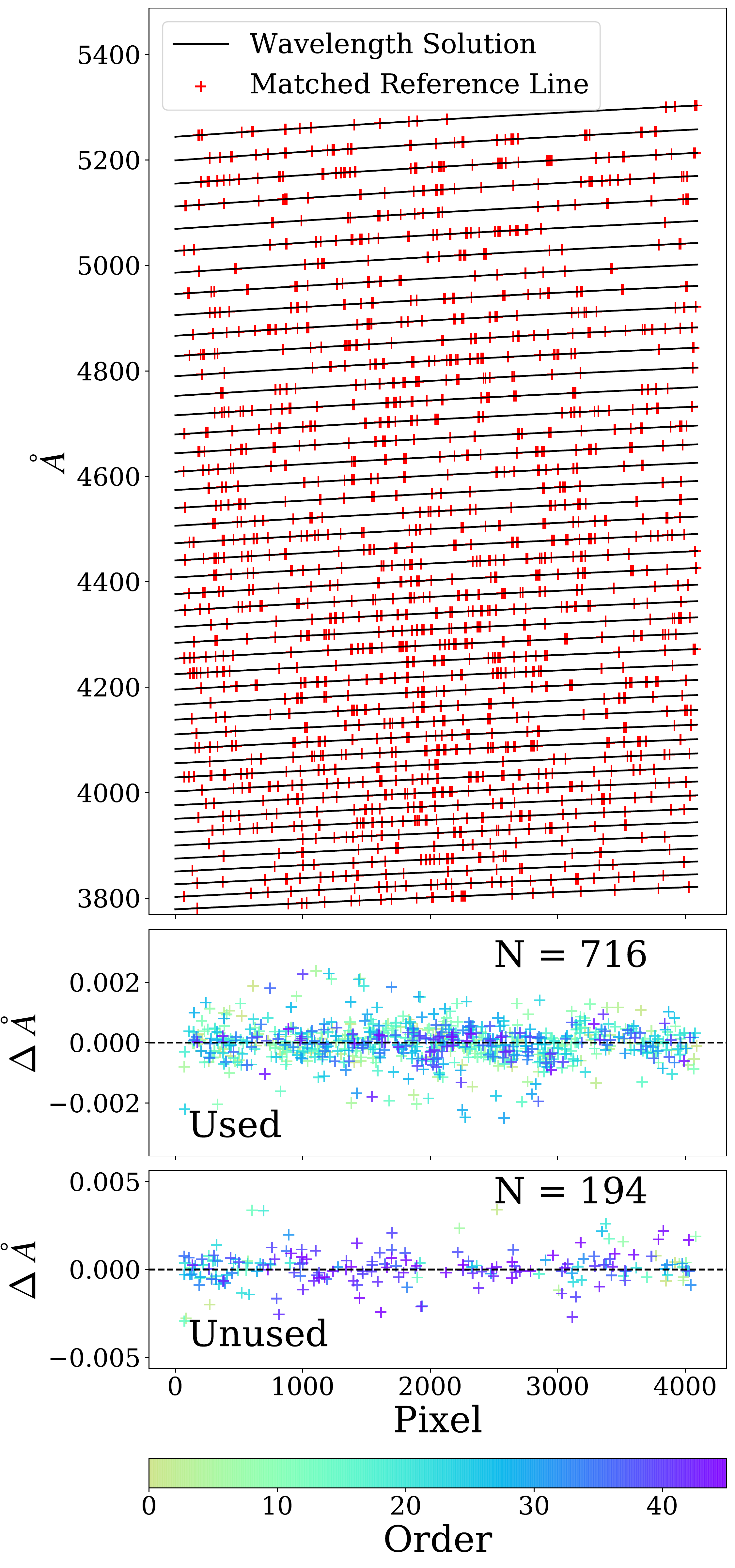}
    \caption{Left: NRES residuals from the Sutherland, South Africa NRES unit. Right: HARPS residuals from calibrating only the blue-most 45 diffraction orders (orders 116-161). The top panels show the wavelength solutions evaluated at every pixel and order as black lines. The upper most line is the red-most order and the bottom-most line is the blue-most order. Each red hatch mark gives the pixel position of an emission-line and the wavelength of its closest-match in the reference wavelength list (the \textit{ESO ThAr atlas} here). Middle panel: the residuals between the ThAr lines which constrained the wavelength solution and their closest matches in the reference list, color coded by diffraction order. Bottom Panels: the low signal-to-noise lines which we did not use to constrain the wavelength solution (our validation set).
    }
    \label{fig:wcs_error_pixel}
\end{figure*}

The residuals per pixel between calibration line wavelengths and their matches in the reference list are shown in Figure \ref{fig:wcs_error_pixel} for the NRES spectrograph at Sutherland, South Africa (left panels) and HARPS (right panels). The top panel shows the wavelength solution evaluated at every pixel and order in black, with each emission line overlaid as a red-hatch. The wavelength solution (black) evolves smoothly as a function of order for both NRES and HARPS. The middle panel shows the residuals of the high signal-to-noise lines which we used to constrain the wavelength solution. Each residual is color coded according to the diffraction order from which the emission line originated. The bottom panel shows low signal-to-noise (S/N between 1 and 3) lines, which we did not use to constrain the solution. We use these lines to cross validate our solution. Those lines have errors comparable to the lines used in the solution, showing that we have not appreciably over-fit the data. 
From the scatter in wavelength space, we estimate the velocity-equivalent precision for HARPS and NRES by weighting all lines equally (after excluding outliers), assuming zero covariance, and assuming that the residuals are Gaussian.

For NRES, the scatter of the 1500 used+unused lines is approximately $\finalscatter$ \AA. 400 lines have errors less than $10^{-3}$ \AA. We calculate a velocity-equivalent precision of \finalprecision \, m/s from the $\finalscatter$ \AA \, scatter. The NRES photon-noise limited precision for the ThAr exposures used in this work is\footnote{from $c \cdot 1/\text{SNR} \cdot 1/\sqrt{N} \cdot 1/R$ assuming $N=3000$ calibration features with an average signal-to-noise per feature of $\text{SNR} = 50$ and resolving power $R = 50000$} $\sim$2 m/s. 

The \texttt{xwavecal} calibration of HARPS yielded smaller residuals compared to NRES. The 900 used+unused lines have a scatter of $5 \cdot 10^{-4}$ \AA, which is a 1 m/s velocity-equivalent precision. This agrees with the instrument's reported velocity precision of 0.90 m/s \citep{HARPSm0}. Additionally, a larger fraction of HARPS lines were close matches: roughly 700 lines (middle panel) constrained the solution while the 200 in the bottom panel did not. This is likely because the \textit{ESO ThAr atlas} is designed for instruments with HARPS' resolving power. Weighting the features by their inverse variance (variance in the pixel centroid position) prior to solving did not reduce the scatter in the residuals for either instrument.

\subsection{Synthetic Data}\label{sec:synthetic_data}
We test our algorithm using synthetic data that consist of two parts: a list of pixel and order positions of spectral features, and a reference list of wavelengths. These data are what one would have after extracting and centroiding the spectral features. We first create a random reference wavelength list with $m$ wavelengths: $\lambda_k$, $k = 1, \ldots, m$. We then map that wavelength list onto pixel and order positions by using the $\lambda(x,i)$ that we found in practice for NRES. Specifically, we apply the inverse of the wavelength solution to map that list into pixel and order positions $(x_j, i_j)$, $j = 1, \ldots, n$ of $n$ spectral features. Note $n > m$ because of the overlaps. We modelled our synthetic data after NRES so that the data are realistic.

For perfect data, every feature position $(x_j, i_j)$ would map exactly to a wavelength $\lambda_k$ in the reference list. We add and omit features to generate more realistic and challenging data. We calibrate synthetic spectra that contain three types of features: those from a line list known to the algorithm $N_{\text{in list}}$, features missing from the reference list $N_{\text{other}}$ (e.g.~contamination from trace elements), and bad features $N_{\text{bad}}$ (e.g.~cosmic rays or bleed-over from saturated pixels in an adjacent order). 

Note that `bad' features are neither in the line list nor are repeated in the overlaps. In real data, these would come from cosmic ray hits, poor centroiding (e.g.~line blends), calibration artifacts, or features that were identified in one side of the overlap but not the other. A good line detection algorithm on well reduced data should not interpret many cosmic ray hits as valid spectral features and the number of `bad lines' should be small.

We calibrate synthetic data with large numbers of bad and missing features to test the robustness, accuracy, and precision of our algorithm. We define a correctly identified spectral feature as one whose inferred wavelength, $\lambda(x,i)$, is closer to the spectral feature's laboratory wavelength than to any other feature's inferred wavelength. This corresponds to an error no larger than the smallest gap in the reference wavelength list. If every spectral feature satisfies this criterion and outliers have been rejected, then convergence is guaranteed on the next solving iteration. Convergence is guaranteed because the true (unknown) wavelength solution is smooth, i.e., free of large variations on sub-Angstrom scales.

\textit{Precision:} 
For our precision tests, the reference wavelength list is a random sample of 2200 from a distribution uniform between 3000 and 9000 \AA. This produces a reference list statistically similar to the \textit{ESO ThAr atlas}. Each measured spectrum contained 1500 features: 1000 features were drawn at-random from the reference list, 500 features were missing from the reference list, and $N_{bad}=0$. These demographics are near what we estimate for NRES. We describe that estimate in Appendix C.

\begin{figure}[]
    \centering
    \includegraphics[width=\columnwidth]{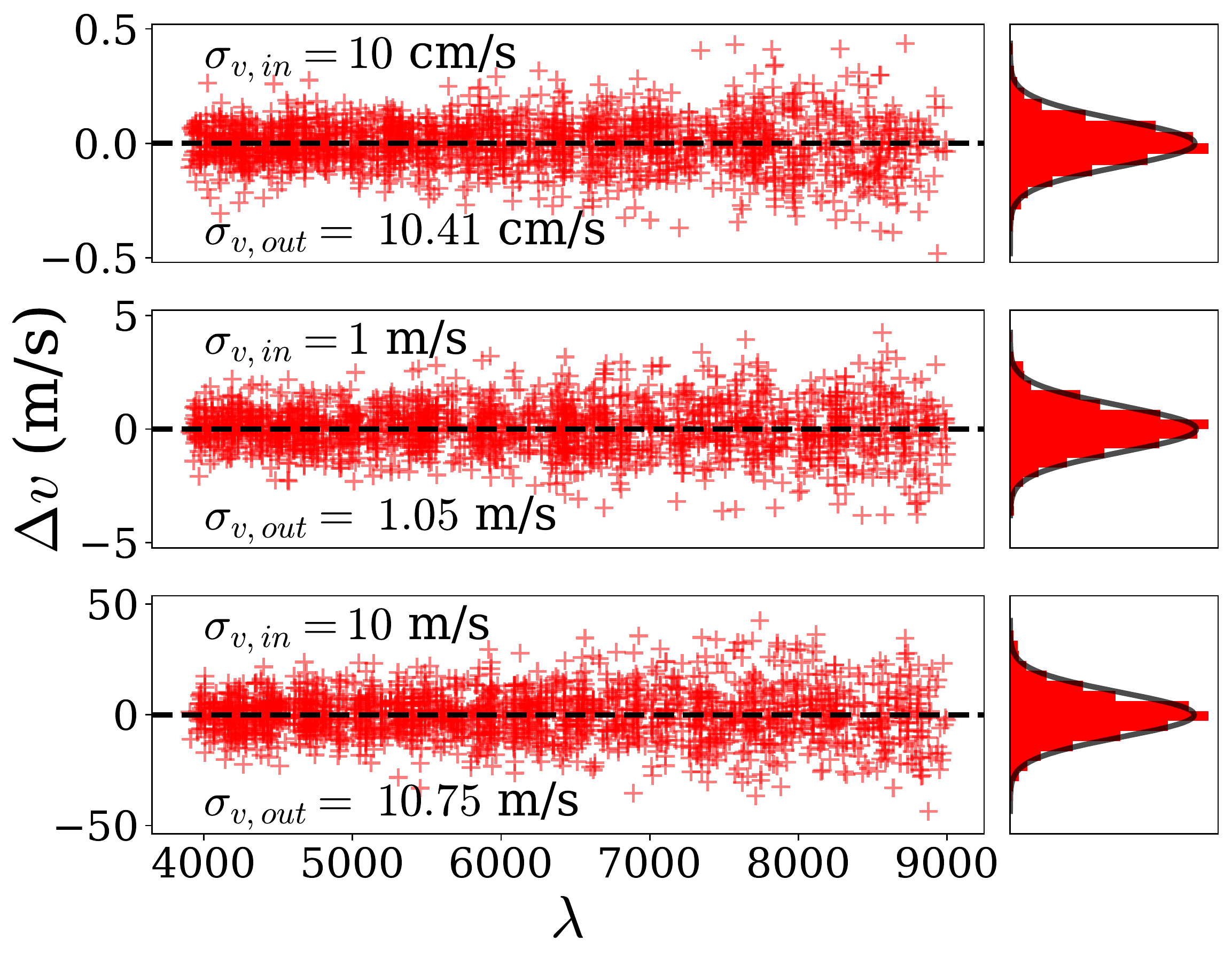}
    \caption{The velocity-equivalent residuals from wavelength calibrating synthetic data which is modelled after NRES, for three levels of velocity-equivalent Gaussian centroid errors ($\sigma_{v, in}$): 10 cm/s, 1 m/s and 10 m/s. In each panel, the injected error $\sigma_{v, \rm in}$ is shown near the top left, and the error estimated from the residuals after calibrating the spectrum, $\sigma_{v, \rm out}$, is shown below it. To the right of each panel, we have plot a Gaussian with zero mean and standard deviation $\sigma_{v, \rm in}$ on top of the histogram of residuals.}
    \label{fig:synthetic_residuals}
\end{figure}

We add Gaussian noise to the pixel positions of each of the 1500 measured features. Given a desired velocity error $\sigma_v$ for each feature in the synthetic data and a known wavelength solution $\lambda(x, i)$, the Gaussian noise that we add in pixel-space to a single feature with wavelength $\lambda_0$ is 
\begin{equation}\label{eq:synthetic_noise}
\sigma_x = \frac{dx}{d \lambda}\frac{\sigma_{\lambda}}{\lambda \sqrt{N}} \lambda_0 = \frac{dx}{d \lambda}\frac{\sigma_v}{c \sqrt{N}} \lambda_0.     
\end{equation}
In Equation \eqref{eq:synthetic_noise}, $N$ is the number of spectral features to be calibrated, $dx/d\lambda$ is the inverse of the first-derivative of the wavelength solution evaluated at $\lambda_0$, and $c$ is the speed of light. Figure \ref{fig:synthetic_residuals} shows the wavelength residuals from calibrating synthetic data with Gaussian noise corresponding to 10 cm/s, 1 m/s, and 10 m/s. The input $\sigma_{v, in}$ and output velocity-equivalent error $\sigma_{v, out}$ is shown in each panel. In all three panels of Figure \ref{fig:synthetic_residuals}, every one of the 1500 features was identified correctly. To the right of each panel, a Gaussian with zero mean and variance $= \sigma_{v, in}^2$ is plotted on top of the residuals histogram. In each case, the velocity-equivalent error $\sigma_{v, out}$ we achieved after wavelength calibration was within 10\% of the input: the systematics of our model and algorithm account for less than 10\% of the error budget for these data.

\begin{figure}[]
    \centering
    \includegraphics[width=\columnwidth]{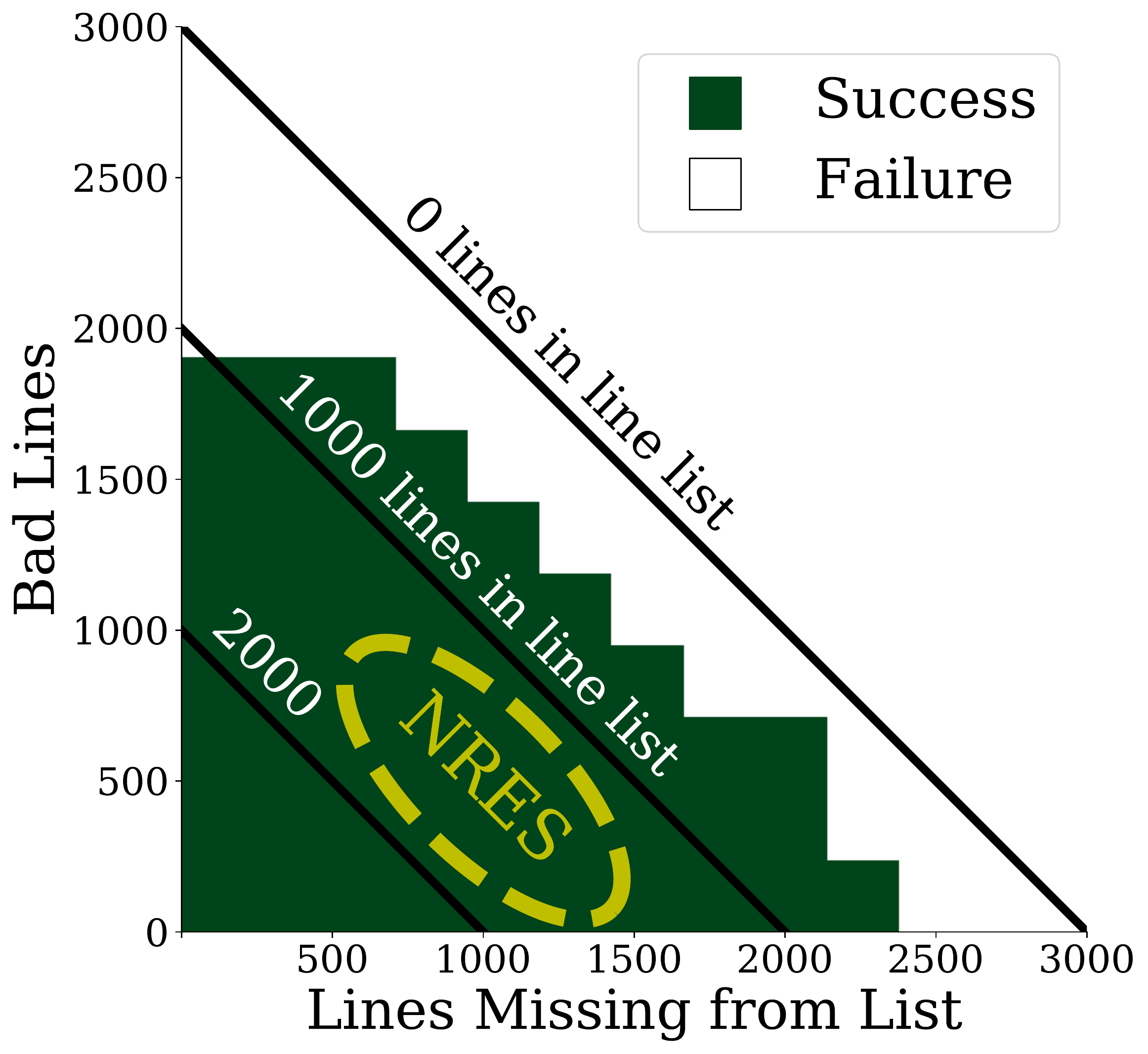}
    \caption{
    Synthetic spectra where our algorithm succeeds (in green) and fails (in white). Each spectrum contains 3000 features, and the reference list fed to our algorithm contains a subset of those features and contamination. We have labelled contours where the reference wavelength lists are of equal length. The region above the solid black line is where the reference wavelength list known to the algorithm is empty. The region within the dashed ellipse is our estimate for the demographics for NRES emission features on a typical calibration exposure (Appendix C).}
    \label{fig:success_rate_with_contam}
\end{figure}

\textit{Robustness:} We calibrate synthetic spectra with varying $N_{\text{bad}}, N_{\text{other}}$ and $N_{\text{in list}}$, subject to the constraint that $N_{\text{in list}} + N_{\text{other}} + N_{\text{bad}} = 3000$. 3000 is roughly the number of features desired in an average-length calibration exposure for an instrument like NRES or HARPS. Figure \ref{fig:success_rate_with_contam} shows what combinations of $N_{\text{bad}}, N_{\text{other}}$ and $N_{\text{in list}}$ lead to successful calibrations. The solid green areas are where the algorithm correctly identifies every one of the 3000 features in the spectrum (that were not bad features). Moreover, such areas have tiny median absolute deviations and so a user would have correctly recognized them as successful. Our algorithm succeeds even when $1800$ features (60 \%) are bad, and when only $20\%$ of the features are in the reference wavelength list. The white regions are where our algorithm incorrectly identifies at least one feature, which we deem a failure. Failures can easily be identified as such based on the median absolute deviation of the residuals alone. Our algorithm begins to fail when bad features constitute more than 60\% of the population, because at that point there are only $\sim$15 features from the lamp in each order and so the $6$ matched feature minimum for our overlap algorithm is never met. Our algorithm is more resilient to features missing from the list (80\% can be missing), because missing features still help fit the overlaps while bad features do not. In Figure \ref{fig:success_rate_with_contam}, we have included the demographic estimate for NRES to place the synthetic data in context (see Appendix C). NRES is well within the region of parameter space where our algorithm succeeds.

\section{Discussion}\label{sec:discussion}
The positions of spectral features for stable instruments like HARPS or NRES drift at-most by fractions of a pixel throughout the night because the optics are fixed to within tens of microns. By contrast, few pixel shifts are commonplace for instruments like NIRSPEC \citep{MartinNIRSPEC, Mclean1998DesignAD} where filters and optics move to accommodate observation requests across many photometric bands. Our algorithm is appropriate for both types of instruments. For very stable instruments, our algorithm could be run in full on each exposure, or used infrequently on the timescale over which the instrument is stable (e.g.~monthly). In the latter case, low-order perturbations would correct for small shifts between nights. For instruments where angstrom shifts are expected between nights or even observations, our algorithm provides reliable absolute calibrations simultaneous with observations.

Our algorithm is designed for absolute calibration sources, such as arc lamps or absorption cells. Sources such as laser frequency combs (LFCs), possessed by e.g. the Infra-red Doppler (IRD) instrument \citep{Kotani+etal_2018_IRD}, or Fabry-Perot (FP) Etalons, emit a finely spaced, equidistant grid of spectral features which can be used to refine a wavelength solution anchored by an absolute calibration lamp such as ThAr. Although the spacing between any two FP or LFC features is known precisely, the exact wavelength of any given line cannot be calculated precisely enough for absolute wavelength calibration. One can use the locations of FP and ThAr lines together to refine the \texttt{xwavecal} wavelength solution, e.g. by using the differences between FP lines as constraints on the first derivative of the wavelength solution. See \cite{Cersullo+Coffinet+etal_2019_HARPSFP} and \cite{Petersburg+etal_EXPRES_2020} for examples of combining FPs or LFCs with absolute calibration sources. We aim to add LFC/FP combined solutions to future versions of \texttt{xwavecal}.

We tested our algorithm on five instruments: HARPS and the four NRES spectrographs. Our wavelength solution on HARPS has an estimated precision comparable to their quoted value of 1 m/s. Our precision on NRES is $\finalprecision$ m/s. This is a factor five above the NRES photon-noise limited precision of ${\sim} 2$ m/s but is sufficiently precise for every line to be identified correctly. Still, this reduced precision could be caused by instrument effects like elongated and skewed line-spread-functions, modal noise, systematics in the reference list, or the lack of rejecting cosmic rays from our feature-list. The \finalprecision \, m/s precision is not likely due to model/algorithm problems as we showed by perfectly calibrating synthetic data modelled after NRES and by reaching 1 m/s on HARPS. In a future publication, we will characterize NRES in detail and show the RV precision that we can achieve.

With realistic synthetic data, we showed that when our algorithm succeeds, it correctly identifies every single spectral feature. Moreover, any successes or failures are identifiable as such from the scatter alone. It correctly calibrates spectra where $60\%$ of the detected features are bad (e.g.~cosmic rays hits). Our algorithm succeeds on spectra even when $80\%$ of the features have no counter-part in the reference list. Although it is possible for a large number of thorium or argon features to be undocumented, $80\%$ contamination is far beyond the level expected in any \'echelle spectrograph and serves only as a stress test for our algorithm. For instance, \cite{2007MNRAS.378..221M} estimate that $<1\%$ of the features on a UVES (Ultra-violet and Visual \'Echelle Spectrograph) calibration frame are from contaminants such as Na, Mg, Ca and Fe.
Our algorithm is robust and reliable given a modest set of overlaps and appropriate wavelength models at each stage. The other parameters are either easy to measure or our algorithm finds them automatically. We now turn to sensitivities.

\textit{Requirements with overlaps:}
Although only $\sim$5 well-fit, closely-spaced overlaps are needed, more overlaps mitigate failures. However, too many (e.g.~$60$) makes the simple model of Equation \eqref{eq:solve_wcs_for_overlap} (which lacks degrees of freedom independent of $x$) insufficient to model the curvature over such a wide wavelength range. The default \texttt{xwavecal} parameters succeeded on NRES and HARPS with $5-30$ overlaps fit across $\sim30$ consecutive orders. Note that we include every order when we refine, only the initial overlap fitting stage is restricted to the $\sim30$ consecutive orders mentioned.

\textit{The reference wavelength list:}
The reference wavelength list can contribute errors at the m/s level. For instance, the wavelengths in the original \textit{ESO ThAr atlas} (which we used in this work) are based on effective wavelengths derived by \cite{deCuyper_line_list}. The effective wavelengths are weighted averages of ThAr lines that would be blended together for a given resolving power $R$. The \textit{ESO ThAr atlas} is designed for spectrographs with $R \approx 10^5$, specifically UVES \citep{2007MNRAS.378..221M}. Most ThAr lines are blended \citep{2007MNRAS.378..221M}, and therefore most of the wavelengths in the list are likely slightly incorrect for NRES which has half of the resolving power for which the line list was designed. The 10 m/s performance on NRES compared to the photon limit of 2 m/s could in part be due to the fact that the \textit{ESO ThAr atlas} is designed for the resolving power of the latter and not the former. Generating the optimal reference wavelength list is a solved problem. \cite{2007MNRAS.378..221M} detail how to generate the optimal reference list for any given instrument, provided a ``reduction code which allows repeatable wavelength calibration with different ThAr line-lists". \texttt{xwavecal} provides repeatable wavelength calibration with arbitrary reference wavelength lists. Anyone can now generate a library of optimal reference lists for a variety of instruments, i.e. implement the \cite{2007MNRAS.378..221M} algorithm and couple it with the \texttt{wavelength\_calibrate} function of \texttt{xwavecal}.

\textit{The wavelength model at each stage:}
The model degrees $N_x$, $N_i$ used at each of the three stages (solving from the overlaps, constraining over the detector, and final refinement) are important. An incorrect number of degrees of freedom at early stages may cause over-fitting and failure. At each stage, one wants $N_x$, $N_i$ sufficiently large to reproduce the complexities of the wavelength solution yet not so large that the model will overfit the data. We discussed our framework for choosing the final $N_x$, $N_i$ in Section \ref{sec:choosing_model}. One could apply the same procedure for the models at earlier stages: keep the final model fixed and minimize the residuals as the other two models are made less complex. We have tested that the \texttt{xwavecal} default models work well for spectrographs with $R$ between 20,000 and 120,000. We recommend that users change the models only in exceptional cases.

\textit{Identifying spectral features:}
The global scale search is sensitive to the number of identified spectral features. The search may fail if too many low signal-to-noise lines are included because many weak lines are either contamination from other elements, e.g.~tantalum \citep{Pakhomov2015}, or are missing from any reference list. If one identifies too few lines, the global scale search may fail because many of the reference wavelengths were not identified, or one may not be able to fit the overlaps due to the lack of spectral features. For NRES, those failure modes occur if more than $\sim$2000 or fewer than $\sim$500 lines are identified (within a single fiber). One can safely add low signal-to-noise lines after final refinement, when the solution is well constrained and resilient to contamination.

\textit{Failed wavelength solutions:}
A successful wavelength solution will have good overlaps between every order: every duplicated feature between two orders in an overlap will be aligned. Because we do not anchor lines, our algorithm fails catastrophically when it fails at all (see Figure \ref{fig:m0_merit}), meaning that the residuals will have a median absolute deviation on the order of the spacing, e.g.~${\sim} 0.5$ \AA, and there will be many orders with poor overlaps (many mis-aligned features). Large residuals are the quickest way to identify a failed wavelength solution, and small residuals across hundreds of spectral features indicates a success.

\section{Conclusions}\label{sec:conclusion}
We have developed an algorithm for finding the wavelength of each pixel on an \'echelle spectrograph via the constraint that duplicated features in a spectrum must map to the same wavelength. This constraint establishes the scale-invariant components of the wavelength solution. We then solve for the scale and higher order components by matching the central wavelengths of spectral features to laboratory wavelengths. We demonstrated our method's precision by calibrating realistic synthetic data to arbitrary precision (cm/s and lower). On synthetic data, our model and algorithm contribute less than $10\%$ to the error budget. We showed our method's robustness by calibrating highly contaminated synthetic data. Our method achieves velocity-equivalent precisions of 1 m/s on HARPS and \finalprecision \, m/s on NRES, assuming that the errors of individual line positions are uncorrelated. The latter is worse than the photon-noise limit of $\sim$2 m/s but could be due to line-profile and line list effects in NRES for which we have yet to correct.  We demonstrated our method's generality and precision-in-practice by calibrating HARPS and reaching the spectrograph's reported precision of  ${\sim}$1 m/s. We provide an open-source \textit{Python} implementation of this algorithm, \texttt{xwavecal}, installable by name via \texttt{pip} and by download at \url{https://github.com/gmbrandt/xwavecal}.

\acknowledgments
{We thank the anonymous referee for a thoughtful and detailed review. We thank Las Cumbres Observatory for supporting this work. G.M.~Brandt thanks Daniel Harbeck (Las Cumbres Observatory) and Craig Pellegrino (University of California Santa Barbara) for insightful and in-depth feedback on early drafts of this work. We thank Emily C.~Martin (University of California Santa Cruz) and Cullen Blake (University of Pennsylvania) for helpful conversations.

Las Cumbres Observatory has received considerable support for NRES through NSF MRI (AST-1229720) and ATI (1407666 \& 1508464) grants.

This work made use of observations from the Las Cumbres Observatory network, and of observations collected by the European Organisation for Astronomical Research in the Southern Hemisphere. 

This research made use of Astropy, a community-developed core Python package for Astronomy \citep{astropy:2013, astropy:2018}.}

\appendix

\section{A. The algorithm for fitting the overlaps}
Given a pair of 1-D spectra from adjacent orders, $f_{i}(x)$ and $f_{i+1}(x)$, let the set of features (e.g. emission lines) in each spectrum be $\{ (x_n, i) \}$, $n = 1,...,N$ and $\{ (x_m, i+1) \}$, $m=1,...M$, respectively; where $(x_n, i)$ implies a spectral feature in order $i$ that is centered at pixel $x_n$.
We consider a coordinate transformation $g_i(x)$ to be correct if there are at least $F$ pairs of spectral features $(x_n, i)$, $(x_m, i+1)$ for which $g_i(x_m) = x_n \pm 1$. This feature matching criterion will exclude valid overlaps (false negatives) fewer than $F$ matched features. For example, it may exclude orders with narrow overlaps near the edges of the detector. Any $g_i(x)$ that yields three or fewer matched features should not be trusted because a quadratic can be fit to any set of three points. False negatives are not an issue provided enough true positives are found. Our objective is to find $g_i(x)$ for as many orders as possible, with few or no false positives (incorrect mappings $g_i(x)$ that nevertheless match at least $F$ spectral features). 

We assume the mapping $g_i(x)$ is a $B^{th}$ order polynomial which has $B+1$ free parameters. We take a set of $B+2$ pixel coordinates $x_n$ and a set of $B+2$ pixel coordinates $x_m$, each of which might be chosen at random. We solve for the $B$ polynomial coefficients via least squares that best maps the $x_n$ to the $x_m$. We then transform all of the $x_m$ via $g(x_m)$ and count the number of spectral features that are matched based on the aforementioned criterion. Our algorithm tries every combination of $B+2$ features from the leftmost $N$ on the blue side of the red order, with every combination of $B$ features from the red-side of the blue order. We take the polynomial coefficients that maximize the number of matched features. This concludes the algorithm. We narrow the possible combinations to make this algorithm computationally feasible.

For a sense of the scale of this search, assume $g_i(x)$ is quadratic ($B = 2$) and that $N=15$. Given 40 features total in the blue order, this search na\"ively requires $_{15}C_4 \times \ _{40}C_4$ trials or roughly 125 million matrix inversions to find all possible sets of 3 polynomial coefficients. This is intractable on a ordinary computer. We reduce the number of trials by trying only monotonic combinations (i.e. only where $x_{m-1} < x_{m}$ and likewise for $x_n$) because $g_i(x)$ must be monotonic. We restrict the search to consider pairs of features whose fluxes are equal within a threshold. This threshold is set by \texttt{flux\_tol} in the \texttt{xwavecal} configuration file. 
In the absence of instrument-induced flux variation in the spectrum, two duplicated features will have identical fluxes (within the noise). However, interferences in the echelle grating cause modulation in the recorded flux. This modulation depends on the blaze angle of the echelle grating and is named the blaze function. This algorithm succeeds with a small \texttt{flux\_tol} threshold if the feature fluxes have been divided by the blaze function, which is called blaze correcting. Section 2.6 of \cite{CERES-2017PASP..129c4002B} offers one way to blaze correct a spectrum. Providing a blaze corrected spectrum to \texttt{xwavecal} allows for smaller thresholds and faster overlap fitting.

The default parameterization in \texttt{xwavecal} works well for NRES, HARPS, and lower dispersion spectrographs. The threshold \texttt{flux\_tol} is 0.2. One should change it to $\gtrsim 0.6$ if the spectrum is not blaze corrected or if the blaze correction is poor. We truncate $g_i(x)$ at $x^2$ (set $B = 2$) and set $N=15$ by default. $B$ would only need to be changed if your spectrograph has a resolving power $R$ many factors larger than $R = 100,000$. The \texttt{xwavecal} default feature-match criterion of $F = 6$ excludes all false positives on NRES and HARPS spectra. $F$ is set by \texttt{min\_num\_matches} and can be changed easily, but should only be changed in exceptional cases by the user. On NRES with the default parameters, this algorithm successfully fits $\sim40$ of the $67$ overlaps possible and marks $\sim30$ as valid.

\section{B. The global scale search}
We define our global goodness-of-fit metric, $\metric$, as
\begin{align}\label{eq:goodness_of_fit}
    \metric = \sum_n \text{min}| \lambda(x_n, i_n) - \lambda_{\rm ref}| ^2 
\end{align}
where $n$ runs over the calibration spectral features, $(x_n, i_n)$ is the coordinate of the $n$\textsuperscript{th} spectral feature, $\lambda_{\rm ref}$ is the wavelength of a line from the reference list, and $\text{min}| \lambda(x_n, i_n) - \lambda_{\rm ref}|$ is the absolute value of the difference in wavelength between a line and its closest match in the reference list. We only consider spectral features $(x_n, i_n)$ from orders $i_n$ such that $i_{\rm lower} \leq i_n \leq i_{\rm upper}$ where $i_{\rm lower}$ and $i_{\rm upper}$ are the order indices of the red-most and blue-most orders whose overlaps were successfully fit. In other words, we only consider features who lie within the spectral region most well constrained by the overlaps. We introduce $K$ dependence into our goodness-of-fit metric (Equation \eqref{eq:goodness_of_fit}):
\begin{align}\label{eq:goodness_of_fit_bf}
    \metric(K) = \sum_n \text{min}|  K \cdot \frac{\lambda(x_n, i_n)}{K_{0}} - \lambda_{\rm ref} | ^2 .
\end{align}
where $\lambda(x_n, i_n) / K_{0}$ is our dimensionless wavelength solution that we solved for with the overlaps. $K_0$ is the unknown global scale. Equation \eqref{eq:goodness_of_fit_bf} will have a minimum when $K \approx K_0$. 

If the spectrograph is poorly characterized (e.g. during instrument testing), then one may want to search over a large range of $K$ values, e.g.~between $10$ times and $1/10$ times the initial guess. In such cases, the background signal of $\metric(K)$ from randomly matching features in the spectrum and the reference wavelength list dominates and $K_0$ is no longer a global minimum, but instead is the sharpest local minimum. \texttt{xwavecal} implements the following procedure to find the sharpest local minimum.

The expected value of $\metric(K)$ (the random-match continuum) can be calculated from the kernel density estimates of the reference list and identified features. \texttt{xwavecal} uses the median filtered $\metric(K)$ as a computationally cheap replacement for the random-match continuum. We divide $\metric(K)$ by the continuum since we are looking for a local decrease in the fraction of features with bad matches. We conclude the brute force search by taking the $K$ value that yields the global minimum of the continuum divided $\metric(K)$. We then refine $K$ by minimizing $\metric(K)$ with a standard optimizer.

\begin{figure}[t!]
    \centering
    \includegraphics[width=0.7\columnwidth]{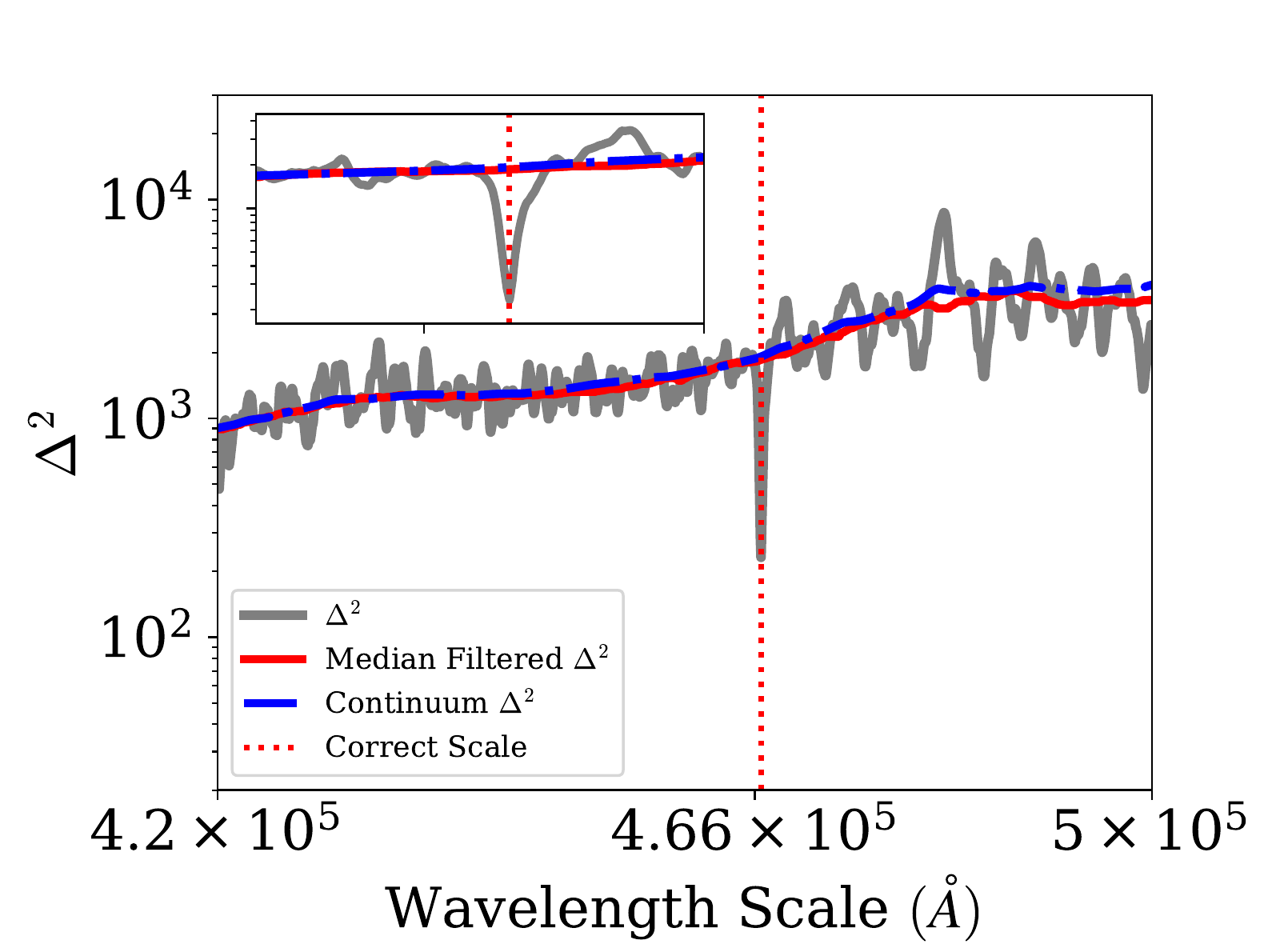}
    \caption{In grey is the $\metric(K)$ evaluated over a small range in $K$ for the NRES spectrograph at Sutherland, South Africa. The true global scale $K$ yields the sharp local minimum near $K = 4.66 \cdot 10^5$ \AA \, and is highlighted by the dashed red line. The continuum expected from randomly matching features in the spectrum with features in the reference list is shown by the blue dash-dot curve, and the median filtered signal, which closely approximates the continuum, is shown in red.}
    \label{fig:chisqvslambdascale}
\end{figure}

Figure \ref{fig:chisqvslambdascale} shows the landscape of $\metric(K)$ on the NRES Sutherland spectrograph. A large global minimum at small $K$ (not shown) and the multitude of local minima in $\metric(K)$ in Figure \ref{fig:chisqvslambdascale} makes the minimization problem difficult and necessitates a brute force approach.

\section{C. Estimating spectral feature demographics}

Here we describe how we estimate the fraction of features on a detector which are: captured in the reference list; not in the reference list yet are real spectral features; or are spurious contamination (e.g.~cosmic rays). We refer to these populations as $N_{\text{in list}}, N_{\text{other}}$ and $N_{\text{bad}}$, respectively.  We assume the instrument has been calibrated with \texttt{xwavecal} and so quantities like the number of matched features across all overlaps are known. We assume $N_{\text{in list}}$ is equal to the average number of close matches in a wavelength solution (e.g.$\sim$1500 for NRES). We now set up two equations which can be solved for $N_{\text{in list}}, N_{\text{other}}$ and $N_{\text{bad}}$.
\begin{align}
    N_{\text{outliers}} &= N_{\text{bad}} + N_{\text{other}} \label{eq:C1} \\
    N_{\text{overlaps}} &= A(N_{\text{in list}} + N_{\text{other}}) \label{eq:C2}
\end{align}
In Equations \eqref{eq:C1} and \eqref{eq:C2}, $N_{\text{outliers}} = N_{\text{total}} - N_{\text{in list}}$ where $N_{\text{total}}$ are the total number of observed features. $N_{\rm overlaps}$ is the sum of the number of matched features across all overlaps. $A$ is a proportionality constant that is the product of the redundancy of the detector (e.g.$\sim$1/2 for NRES) and the ratio of overlaps fit to total number of overlaps on the detector (1/2 to 1/3 for NRES). Equation \eqref{eq:C2} is valid because matched features can only come from real features regardless of whether or not they are in the line list. The sum of the number of matched features across all orders must then be proportional to $(N_{\text{in list}} + N_{\text{other}})$. Equations \eqref{eq:C1} and \eqref{eq:C2} can be readily solved for $N_{\text{bad}}$ and $N_{\text{other}}$. We now calculate these quantities for NRES.

Figure \ref{fig:wcs_error_pixel} has 3000 features total of which $\sim$1500 are close matches. So, $N_{\text{in list}} = 1500$ and $N_{\text{outliers}} = 1500$. For NRES, $A \lesssim 1/5$. Solving Equations \eqref{eq:C1} and \eqref{eq:C2} aforementioned equations with $A \leq 1/5$ gives $N_{\text{other}} \in (400, 1500), N_{\text{bad}} \in (1100, 0), N_{\text{in list}} = 1500$, for 3000 features total.

\bibliographystyle{apj_eprint}
\bibliography{refs}

\end{document}